%% file: main.tex
\documentclass[conference]{IEEEtran}

\usepackage{amsmath,amssymb,amsfonts}
\usepackage{booktabs}
\usepackage{xcolor}
\usepackage{graphicx}
\usepackage{hyperref}
\usepackage{url}
\usepackage{algorithm}
\usepackage{algpseudocode}
\usepackage{subfig}

\newif\ifcomment
\commentfalse
\ifcomment
\newcommand{\sadikshya}[1]{{\bf \textcolor{purple}{Sadikshya: #1}}}
\newcommand{\jaishnoor}[1]{{\bf \textcolor{blue}{Jaishnoor: #1}}}
\newcommand{\taylor}[1]{{\bf \textcolor{orange}{Taylor: #1}}}
\newcommand{\josef}[1]{{\bf \textcolor{green}{Josef: #1}}}
\newcommand{\sayak}[1]{{\bf \textcolor{red}{Sayak: #1}}}

\else
\newcommand{\sadikshya}[1]{}
\newcommand{\jaishnoor}[1]{}
\newcommand{\taylor}[1]{}
\newcommand{\josef}[1]{}
\newcommand{\sayak}[1]{}
\fi

\ifCLASSINFOpdf
\else
\fi
\begin{document}
\title{Binge, Bot, Repeat: Unpacking the Ecosystem of Video Piracy on Telegram}


\author{\IEEEauthorblockN{Sadikshya Gyawali}
	\IEEEauthorblockA{Louisiana State University\\
		sgyawa5@lsu.edu}
	\and
	\IEEEauthorblockN{Jaishnoor Kaur}
	\IEEEauthorblockA{Louisiana State University\\
		Jaishnoor.Kaur@lsu.edu}
        \and
	\IEEEauthorblockN{Taylor Graham}
	\IEEEauthorblockA{Louisiana State University\\
		tgrah22@lsu.edu}
                \and
	\IEEEauthorblockN{Josef Horacek}
	\IEEEauthorblockA{Louisiana State University\\
		jhoracek@lsu.edu}
        \and
	\IEEEauthorblockN{Nowshin Tabassum}
	\IEEEauthorblockA{University of Texas at Arlington\\
		nxt6577@mavs.uta.edu}
        \and
	\IEEEauthorblockN{Shirin Nilizadeh}
	\IEEEauthorblockA{University of Texas at Arlington\\
		shirin.nilizadeh@uta.edu}
        \and
	\IEEEauthorblockN{Sayak Saha Roy}
	\IEEEauthorblockA{Louisiana State University\\
		sayak.saharoy@lsu.edu}
}

\maketitle

\begin{abstract}
Telegram has emerged as a major platform for large-scale video piracy, where copyrighted content is rapidly distributed among users. Despite its prominence, the structural and operational dynamics of this ecosystem remain insufficiently understood. 
To address this gap, we present the first large-scale study of video piracy on Telegram through a mixed-method analysis of 1,057 channels that shared 209k unique posts between December 2023 and January 2026---systematically characterizing their content, distribution strategies, and how the ecosystem is sustained at scale. 

Central to our approach is the development of a fine-grained taxonomy that enables a structured understanding of the activity and intent of these channels on a per-post level. The channels collectively distributed 19,033 unique copyrighted titles originating from 175 countries, accumulating over 4.85B unique views and resulting in a lower-bound estimated financial loss of \$17.49B for content rights holders. We also find that this ecosystem is deliberately engineered to be resilient against takedown efforts, frequently redirecting users through chains of intermediary channels and automated bots that collectively handle hosting, access control, monetization, and channel discovery.

The scale and persistence of this ecosystem motivated the development of \emph{Anti-RIP}, a real-time framework for detecting emerging video piracy communities on Telegram. Anti-RIP utilizes our taxonomy to generate contextual, interpretable insights that stakeholders confirmed improve the triaging action against reported posts and channels. Over a 61-day period, the framework facilitated the takedown of 524 previously unknown piracy channels and 71 bots. To support reproducibility and future research, we open-source both the dataset and the \emph{Anti-RIP} framework.
\end{abstract}


% \fi
\IEEEpeerreviewmaketitle

\input{sections/introduction}

\input{sections/related_work}

\input{sections/qualitative_analysis}
\input{sections/taxonomy}
\input{sections/quantitative_analysis}
\input{sections/realworld_impact}

\input{sections/conclusion}

\bibliographystyle{IEEEtranS}
\bibliography{refs}

\appendices

\input{sections/appendix}
\end{document}

%% file: sections/introduction.tex
\section{Introduction}
Unauthorized online distribution of copyrighted video content, including feature films and serialized television shows, has long impacted the entertainment industry -  causing millions of dollars in losses annually to content creators, distributors, and production studios alike~\cite{ICC2011CounterfeitingPiracyImpacts, NCTA2019PiracyEconomyImpact}. Such attacks were traditionally carried out via peer-to-peer file sharing networks and dedicated streaming websites that allowed users to easily discover, access, and consume pirated content at scale~\cite{mckenzie202025, mooney2010napster, POUWELSE2008701, stout2025siteblocking}. However, the dependence of these systems on identifiable hosting infrastructure made them vulnerable to coordinated legal intervention, leading to rapid domain seizures \cite{fedsSeizeDomainNames2010, usassistsbulgaria} and platform shutdowns \cite{alexander2002peer}.

While effective, these enforcement efforts have also driven the evolution of more robust piracy ecosystems that offer greater resilience, scalability, and accessibility. With over 1 billion monthly active users~\cite{mehta2025telegram1b}, Telegram~\cite{telegram} has emerged as a frontrunner in this shift \cite{dhanwate2025empowering}. 
However, the structural and operational dynamics of video piracy ecosystems on Telegram remain largely unexplored. In particular, it remains unclear how such channels are structured, how content is distributed across them, and how the underlying infrastructure is sustained over time. Gaining such insight \textit{is crucial for enabling effective detection and takedown efforts}, and its lack is evident in the largely reactive and ad-hoc responses by security practitioners and stakeholders alike towards such abuse on the platform \cite{maddusila2024copyright, dhanwate2025empowering, BlockX2024TelegramPiracyPlatform, DMCAForceTelegramDMCA}.

To address this gap, we present the first large-scale analysis of video piracy on Telegram through a mixed-method study of 1,057 channels dedicated to such activity between December 2023 and January 2026. Our approach begins with an in-depth qualitative analysis of 74 channels (Section~\ref{qualitative_analysis}) to comprehensively characterize the ecosystem on a per-post level. Insights from this analysis then inform the development of our taxonomy (Section~\ref{taxonomy}), which automates the inference of post behavior, \textit{providing a structured framework for understanding channel activity}. We then use this taxonomy to quantitatively analyze a much larger selection of 209k posts shared in 983 unique channels (Section~\ref{quantitative_analysis}). This analysis enables us to generalize our qualitative insights at scale, identify the volume of copyrighted content, estimate the financial damage across global regions, and map how channels and bots interact among themselves to make content delivery seamless as well as resilient to takedown.

Our findings further enable the development of \textit{Anti-RIP}, a real-time, open-source framework for identifying and reporting emerging piracy communities on the platform (Section~\ref{real_word_impact}). Anti-RIP utilizes our taxonomy to generate fine-grained, interpretable insights into channel activity, along with supporting artifacts, producing \textit{actionable, evidence-backed reports} that significantly reduce the effort required for stakeholders to verify and respond to piracy activity. Given the substantial financial and intellectual harm caused by large-scale piracy on Telegram, we release both the Anti-RIP framework and the associated dataset at
\href{https://github.com/Scalable-Security-Research-Lab/BingeBotRepeat}
{\color{blue} https://github.com/Scalable-Security-Research-Lab/BingeBotRepeat}, to support further research and enable practical intervention efforts.

%% file: sections/related_work.tex
\section{Related Work}

To address the scale of online video piracy, legal frameworks such as the Digital Millennium Copyright Act (DMCA)~\cite{dmca} and Notice-and-Takedown (N\&TD)~\cite{ntd} enable rights holders to enforce the removal of infringing content. However, these mechanisms rely heavily on the stakeholders’ ability to identify and report violations at scale, with early efforts being manually driven and on a case-by-case basis~\cite{usco2020section512, robotakedown, harris2015reboot}, which has had limited impact on the overall availability of the respective pirated content~\cite{urban2017notice}, as cybercriminals frequently recover through rapid re-uploads \cite{robotakedown, ustr2025review, whacamole_ohagan2024} and mirror sites~\cite{ustr2025review}.

Thus, to improve enforcement scalability, both academia and industry have developed several automated detection techniques based on heuristics, network-level signals, and machine learning. For example, Li et al.~\cite{li2022fighting} proposed a system for detecting pirated video websites by leveraging both content and infrastructure-level features, including domain characteristics and webpage structure. Similarly, Zhang et al.~\cite{zhang2022node} model piracy ecosystems as heterogeneous graphs over shared infrastructure such as IP addresses and third-party services, enabling detection through relational patterns. Organizations such as Google have also developed and implemented proprietary tools to identify such content at scale~\cite{google2016fights}. 

In practice, these approaches enable platform-level interventions such as domain blocking~\cite{stout2025siteblocking}, search engine de-indexing \cite{ntd}, and coordinated takedown operations against major piracy sites~\cite{ piratebay2014disruption, takedown123movues, doj2026hackerforum}.
However, increased enforcement pressure on traditional web-based ecosystems has driven piracy operators to migrate toward alternative platforms. Recent reports highlight the growing role of social media and messaging applications in hosting and distributing pirated content~\cite{euipo2021socialmedia, ustr2025notorious}.
In these environments, existing detection tools have been found to be less effective \cite{dhanwate2025empowering, roy2025darkgram, maddusila2024copyright}, with pirated content often persisting and being redistributed over extended periods~\cite{danaher2020effect, roy2025darkgram}. This resilience stems from fundamental differences in how piracy operates on these platforms, where distribution is dynamic, interaction-driven \cite{ morgia2021darktelegram, bal2023telegrampiracy, lymishchenko}, and less reliant on static, publicly accessible infrastructure.

Addressing this challenge requires a deeper understanding of how piracy is organized, distributed, and sustained within these ecosystems. While prior work has covered piracy on platforms such as Facebook \cite{POUWELSE2008701}, YouTube~\cite{kaye2021copyright}, and Telegram~\cite{roy2025darkgram, maddusila2024copyright}, these studies primarily focus on volume and reach, offering limited insight into the structural and behavioral patterns needed to design effective detection frameworks. For example, Roy et al.~\cite{roy2025darkgram}, in a broader study of Telegram-based abuse, analyze a small sample of 360 posts from 36 piracy channels, (incorrectly) assuming the ecosystem to be predominantly dominated by a direct-download distribution model. While this provides an initial perspective, it captures only a narrow slice of the ecosystem and overlooks several key content signals and operational strategies that we uncover in this work.

To address this gap, we focus specifically on Telegram, which has emerged as a major platform for large-scale video piracy distribution \cite{bal2023telegrampiracy, maddusila2024copyright, dhanwate2025empowering}, conducting not only an in-depth characterization of the ecosystem, but also building a taxonomy that can map fine-grained insights automatically to channels and posts that participate in video piracy on the platform. This in turn also enabled us to build Anti-RIP, a lightweight, open-source detection framework that enables stakeholders to identify, understand, and disrupt video piracy on the platform at scale.

%% file: sections/qualitative_analysis.tex
\section{Exploratory Analysis of Telegram Piracy Channels}
\label{qualitative_analysis}

Characterizing the video piracy ecosystem on Telegram requires understanding it \textit{in-depth}. Thus, we dedicate this section towards qualitatively analyzing a sample of Telegram channels that are engaged in video piracy by carefully examining their posts to identify the defining structural and operational characteristics of the ecosystem. We apply the insights gained from this section towards building a structured taxonomy (Section~\ref{taxonomy}) of video piracy activity on the platform that can be used to automatically capture the fine-grained patterns on a per-post level.

\subsection{Seed Channel Identification}
\label{seed_channel_identification}
Our first challenge was identifying a set of channels engaged in video piracy. We adopted the approach of Roy et al.~\cite{roy2025darkgram} and sourced a catalog of 475 available channels from the "Television" and "TV-Shows" section of Telemetrio \cite{telemetr}, a third-party Telegram channel aggregator. Channels indexed by Telemetrio are typically well-established and popular communities, often characterized by high traffic, large subscriber bases, and sustained activity over time. While a realistic piracy ecosystem likely contains a mix of both emerging and long-lived channels, these highly visible channels provide an ideal starting point for qualitative characterization, as they offer richer signals into mature ecosystem behaviors and infrastructure.

To determine which of these channels were dedicated to piracy activity, we manually examined each channel’s name, description, and 20 most recent posts using the official Telegram application \cite{telegram}, after which two coders independently evaluated whether the channel was engaged in piracy. Coder 1 specialized in computer security, whereas Coder 2 had good experience in computer security and social-media analysis through research and academic coursework. The coders used criteria aligned with the Federal Bureau of Investigation Internet Crime Report 2023 \cite{FBIIC32023Report}, initially identifying piracy channels based on whether posts directly mentioned a movie or TV show and provided a link to directly download it. Using this process, we identified 74 channels dedicated to piracy activity. While this heuristic served us well in building this seed dataset of piracy channels, through this paper, we find that this does not fully capture the broader range of piracy-related behaviors on Telegram.
\newline
\subsection{Content Metadata}
\label{content_metadata}
\begin{figure}[t]
\centering
  \includegraphics[width=\columnwidth]{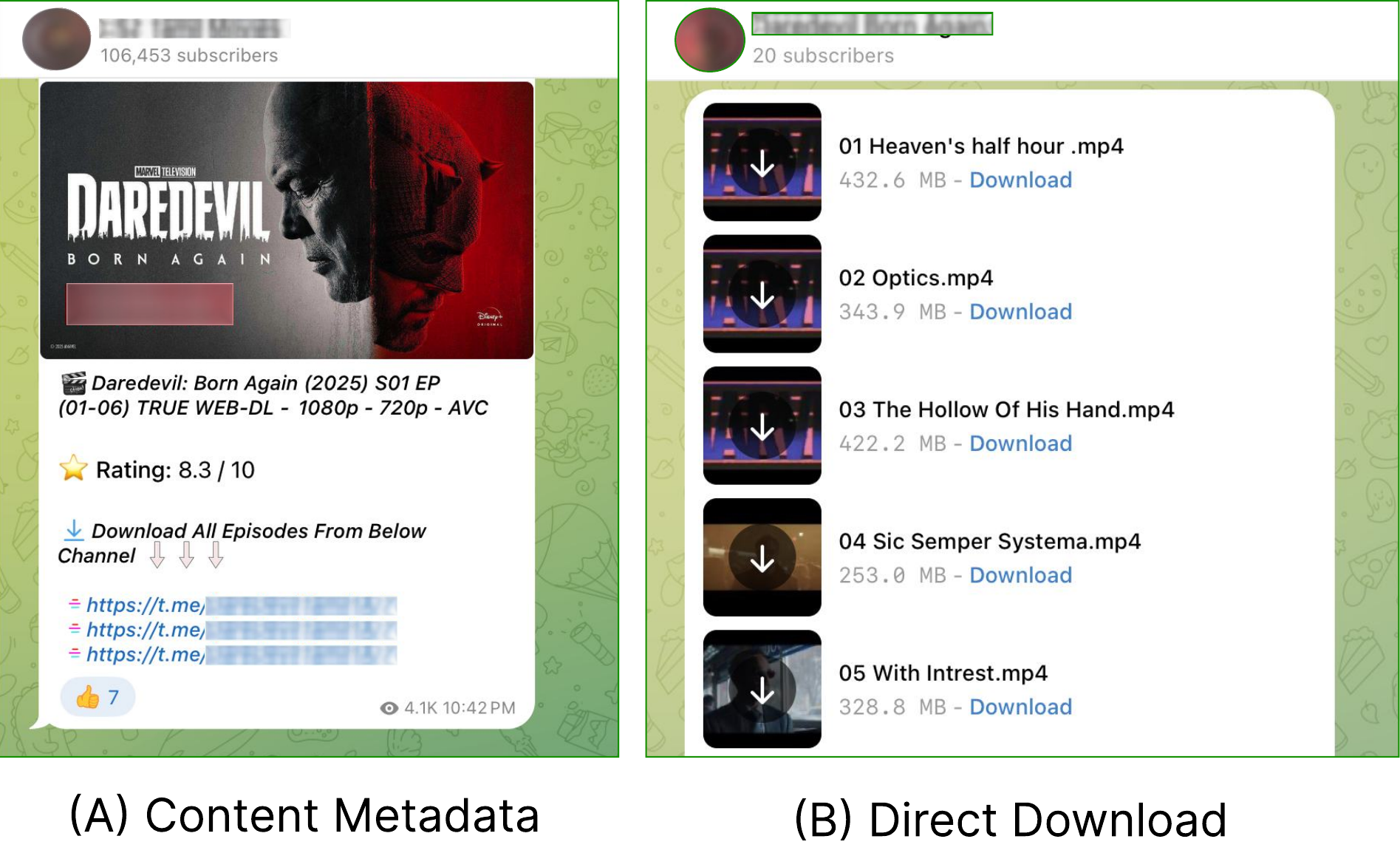}
 \caption{(A) A post sharing a link to download a popular US-based TV show, also containing metadata about the title, such its IMDb rating and video quality. (B) A channel where users can directly download the content.}
  \label{fig:content_metadata_and_direct_download}
\end{figure}

The most prevalent post type across all channels were those which provided access to a specific pirated title.  
These posts, across our seed channels, exhibited a remarkable degree of structural similarity, broadly revealing the recurring characteristics highlighted in the next paragraph. Figure~\ref{fig:content_metadata_and_direct_download} (A) provides an example of the content metadata these posts contain.
\newline
\textbf{Title Metadata:} A large majority of these posts included a poster or banner image of the title, the title name itself, descriptive metadata such as IMDb rating, genre, release year, and language, followed by one or more links through which the content could be accessed. This presentation closely resembles the metadata-rich interfaces of modern piracy streaming websites, suggesting that operators may intentionally emulate familiar content discovery environments. Also, assuming that these channels accumulate large volumes of posts over time, such contextual information helps users search for and distinguish among titles.
\newline
\textbf{Content Quality:} The posts also frequently offered content at multiple resolutions, most commonly 480p, 720p, and 1080p, with 11 channels additionally advertising 4K variants. File sizes were often included alongside each option, allowing subscribers to choose versions based on their interest or bandwidth constraints. While most posts provided access to a single film or individual television episodes, we also observed 112 posts offering either multiple titles within a single post or complete season bundles packaged as unified downloads. In the former case, the grouped films were typically related, such as entries within a film franchise or collections organized around a shared genre or theme. 
\newline
\textbf{Localization:} The majority of these channels (64) focused exclusively on sharing content from a single regional media ecosystem. Among these, 34 channels primarily focused on Hollywood and mainstream American content, 12 on Bollywood releases and Hindi television, 8 on Korean dramas, 7 on Japanese animation (anime), and 4 on South Indian cinema, while additional channels served regional niches including Nollywood, Ethiopian cinema, and Chinese dramas. In contrast, 10 channels functioned as mixed-content aggregators, blending titles across industries and languages. This diversity highlights that Telegram-based piracy is not centered around a single media market but spans a broad international ecosystem of content providers and audiences.
% \begin{figure}[t]
% \centering
%   \includegraphics[width=0.5\columnwidth]{images/localization_cloud_link_backup_channel.pdf}
%  \caption{(A) A post sharing a link to download a popular US-based TV show, also containing metadata about the title, such it's IMDB rating, video quality. (B) A channel where users can directly download the content.}
%   \label{fig:localization_cloud_link_backup_channel}
% \end{figure}
\newline
\textbf{Accessibility:} In 52 channels, posts also offered subtitles and dubbed variants of the same title, a pattern particularly prevalent in Hollywood-focused and Anime channels, where content was made available with subtitles and dubbed audio in languages such as English, Hindi, Tamil, Telugu, Spanish, and Arabic. Figure~\ref{fig:sub_dub_backup_channels} (Left) shows one such example where a title from Japan is provided with English subtitles. In many cases, these variants were presented as separate access options within the same post, allowing subscribers to choose based on language preference. This suggests that operators actively adapt content for broader accessibility, a feature often less consistently supported in traditional piracy websites or peer-to-peer distribution networks, which may help these channels appeal to broader audiences and attract larger subscriber bases.

\textbf{Distribution:} Alongside information about the content itself, these posts typically included one or more links through which subscribers could access or download the material. The destinations of these links varied substantially: some invoked the download directly from the channel, while others redirected users to different channels, automated bots, or external websites. We dedicate the next section to characterizing these distribution strategies in detail.
\subsection{Distribution Strategies}
\label{distribution_strategies}
\subsubsection{Direct File Download}
\label{direct-download-qualitative}
In 579 posts, operators provided direct file downloads hosted entirely within Telegram itself. This is enabled by Telegram’s support for file uploads of up to 2GB per post~\cite{Ashwin2020Telegram2GB}), capacities substantial enough to host several pirated videos directly on the platform. Figure~\ref{fig:content_metadata_and_direct_download} (B) shows one such example.

Allowing operators to host content directly on Telegram provides a significant advantage over conventional piracy websites, which often depend on third-party infrastructure such as file-sharing services to store and deliver content. These third-party services may be monitored, unreliable, or subject to takedown \cite{danaher2014gone}. For example, the shutdown of Megaupload disrupted numerous piracy communities that relied on it as backend storage, instantly breaking access to large volumes of shared content~\cite{danaher2014gone, Knibbs2015RapidShareShutdown}. By hosting content directly within Telegram, operators reduce this dependency and the fragility associated with it. 
\begin{figure}[t]
\centering
  \includegraphics[width=\columnwidth]{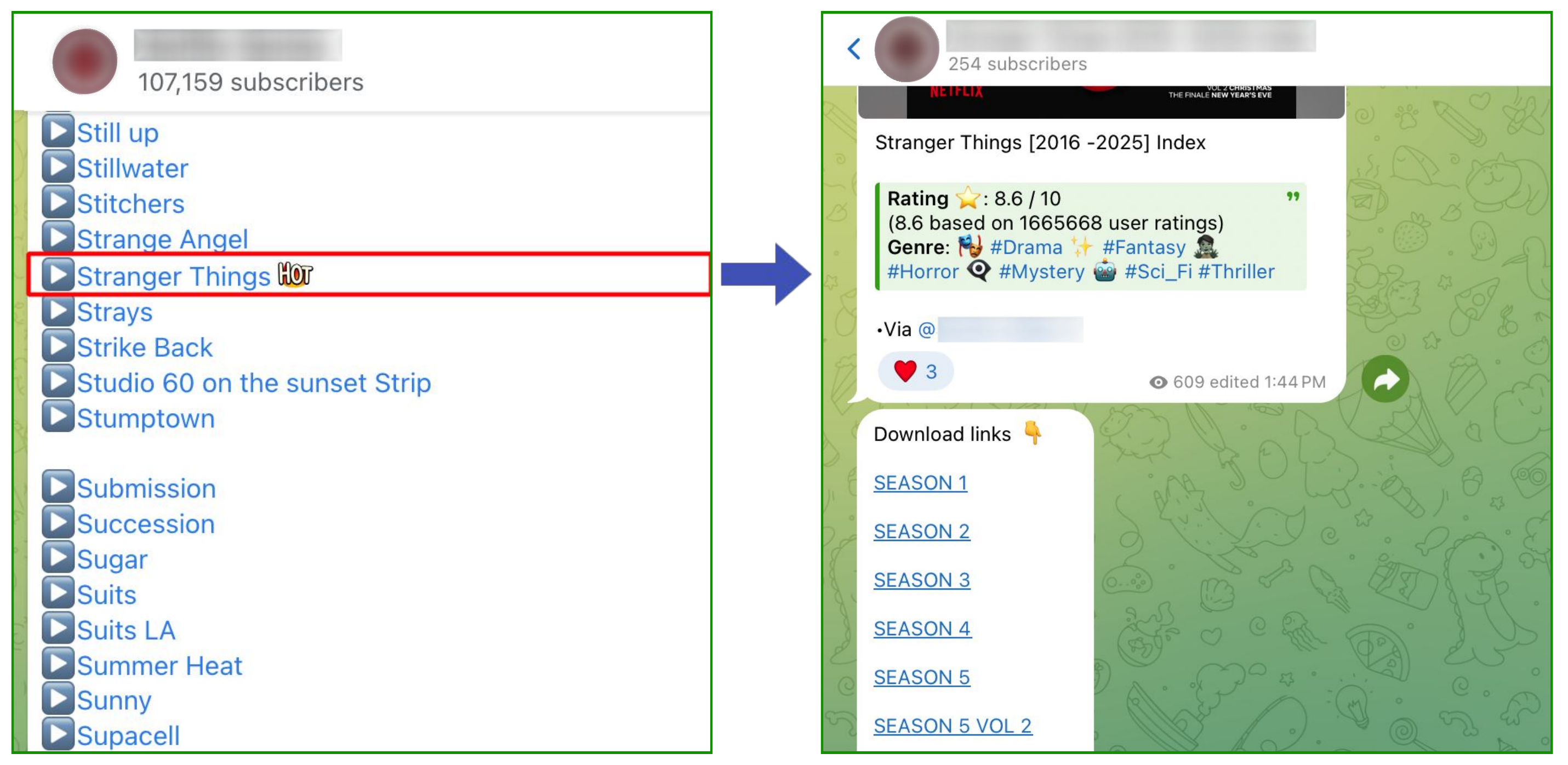}
 \caption{(Left) A channel with an alphabetical index, where clicking on a title leads to (Right), which is a channel exclusively dedicated towards distributing that title.}
  \label{fig:channel_index}
\end{figure}
\subsubsection{Multi-channel Distribution} 
\label{multi-channel-qualitative}
In 642 posts, we observed that operators did not distribute the downloadable file directly within the original post. Instead, these posts redirected subscribers to secondary Telegram channels that either hosted the file themselves or further routed users to external websites where the content could be obtained. Figure~\ref{fig:content_metadata_and_direct_download} presents one such example, where a post advertising a U.S.-based television show redirects users to another Telegram channel that ultimately distributes the downloadable content.

\begin{figure}[t]
\centering
  \includegraphics[width=0.95\columnwidth]{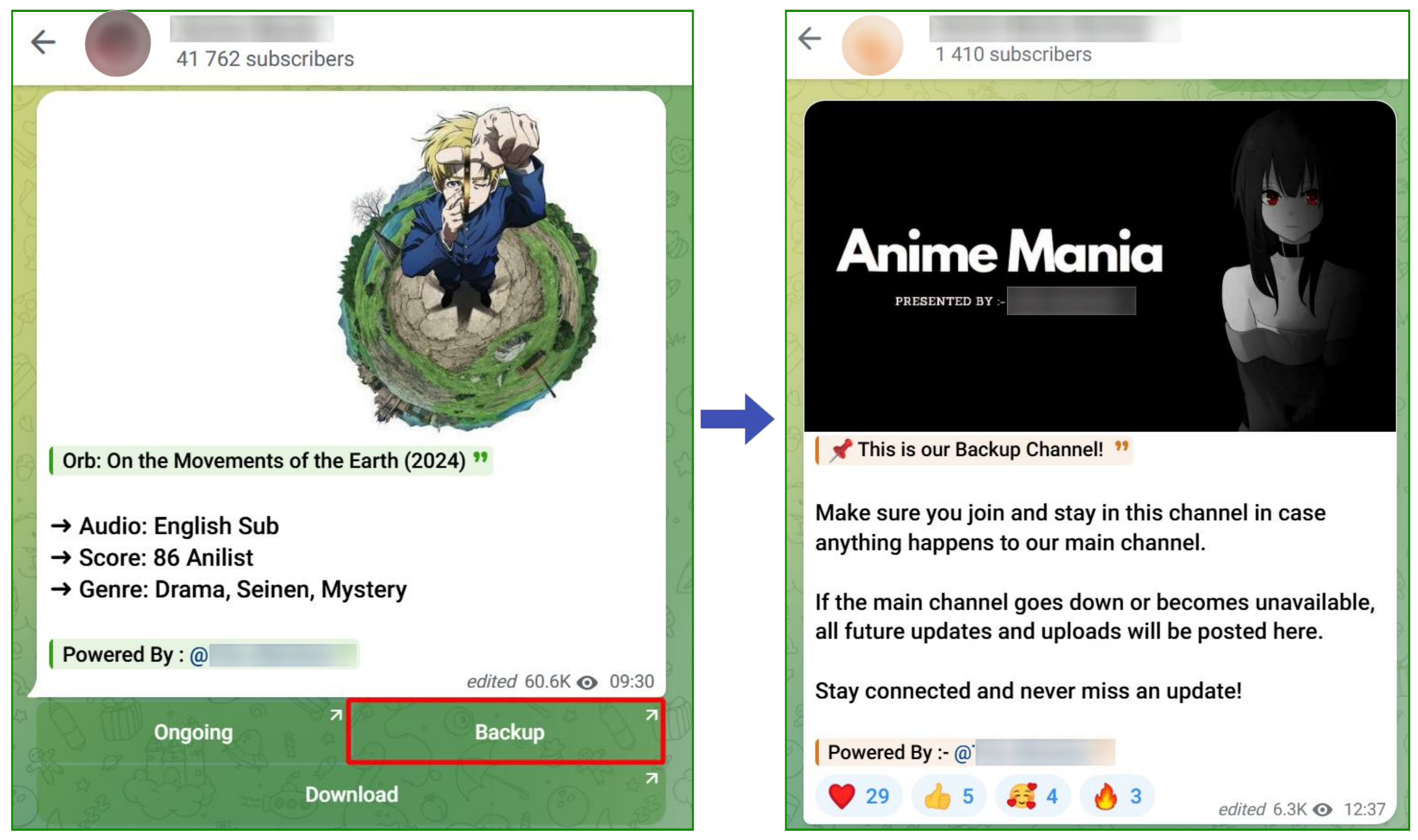}
 \caption{(Left) A channel providing a title from Japan with English subtitles, and a link to a backup channel. (Right) A backup channel which invites users to subscribe to it, in case the main channel gets removed.}
  \label{fig:sub_dub_backup_channels}
\end{figure}

Although the intended purpose of many of these redirections could be inferred from the post text, a substantial fraction of the links (184 out of 642) were no longer functional at the time of analysis. Channels indexed through Telemetrio are generally more visible and therefore subject to greater scrutiny and enforcement pressure than smaller or less discoverable channels~\cite{roy2025darkgram}.  Among the remaining 458 active links, we identified 37 instances where the redirected destination channel was dedicated exclusively to a single title. Figure~\ref{fig:content_metadata_and_direct_download} (B) illustrates one such example, where an entire Telegram channel is used solely to host all episodes of a specific television series.
We further identified 128 cases involving \emph{multi-stage redirection}, where users were routed through multiple intermediary channels before reaching the final download link or external hosting resource. Among these intermediary nodes, 37 channels functioned as directory-like indexing hubs.  Figure~\ref{fig:channel_index} shows one such example, where a channel maintains an alphabetical catalog of titles; selecting a title redirects the user to a separate channel dedicated to that specific content. Additionally, we observed 54 posts where operators explicitly advertised a separate “backup channel” alongside the primary download link. These backup channels specifically communicated to the users that they should subscribe to them in case the original channels get removed. Figure~\ref{fig:sub_dub_backup_channels} illustrates one such example.

Collectively, these behaviors indicate that operators intentionally fragment distribution across multiple interconnected channels rather than centralizing content within a single location. Although this indirection introduces additional steps between the subscriber and the actual content, it also complicates moderation and attribution efforts. Many abuse detection pipelines primarily analyze the initial entry point and do not systematically traverse multi-hop redirections across channels and external websites ~\cite{oest2019phishfarm,bitaab2020scam}, often due to scalability, timing, or resource constraints~\cite{oest2020sunrise}. Subsequently, by isolating content into separate channels or maintaining backup channels, operators reduce the impact of individual channel takedowns, as the removal of one channel can be quickly replaced by another similar channel. 

\subsubsection{External Links}
\label{external_links_qualitative}
Channels also distributed content through one or more external links. To understand how content was delivered through these links, our coders manually clicked and evaluated each of them and analyzed the underlying destination platforms. Overall, they evaluated 728 unique external links across the dataset. Broadly, they found three destinations: cloud storage, streaming websites, and torrent or magnet links.
\newline
\textbf{Cloud Storage:} A total of 412 links pointed to 9 cloud-based file hosting services, where content was uploaded and made available for download through dedicated file-sharing pages. The most commonly observed providers included TeraBox \cite{terabox}, Terashare \cite{terashare}, and GoFile \cite{gofile}. Interestingly, none of these services are considered to be among the major cloud platforms highlighted in prior work \cite{CHAUDHRY201777, Marx2013StorageWars, Makitalo2024TrackingStolenVideo}, where widely used services such as Google Drive, Dropbox and MEGA have been identified as dominant in piracy distribution. This shift suggests that operators in this ecosystem deliberately favor less prominent platforms, likely to reduce visibility and scrutiny associated with mainstream providers. However, our analysis indicates that reduced scrutiny is not the sole motivation. These cloud platforms offer distinct operational advantages that make them particularly attractive to channel operators.

For instance, TeraBox allows free users to upload up to 1TB of data, significantly exceeding the 5–15GB limits typically offered by mainstream cloud providers \cite{AppleSupport108047, GoogleOneSupportStorage}. In contrast, Terashare and GoFile adopt alternative models in which uploads are effectively unrestricted, but access is gated through "credits," or temporary access keys. This shifts the cost burden from the uploader to the downloader, introducing opportunities for monetization within the distribution pipeline, which we discuss later in Section~\ref{monetization}.
Additionally, all three platforms offer paid tiers that provide persistent storage and enhanced access features at a fraction of the cost of comparable plans offered by mainstream providers, in some cases at least 12 times cheaper. 
\newline
\textbf{Streaming Websites and Torrent or Magnet Links:} We identified 49 links directing users to streaming piracy websites, where content could be viewed directly through browser-based interfaces, and 9 links pointing to torrent or magnet-based resources. Compared to direct Telegram uploads and cloud-hosted downloads, the relatively low prevalence of both distribution mechanisms suggests a broader shift away from traditional piracy infrastructures toward platform-based distribution ecosystems. Streaming piracy websites require substantially greater operational overhead, including maintaining hosting infrastructure~\cite{lauinger2013clickonomics} and are more vulnerable to centralized enforcement actions. 
Similarly, although peer-to-peer (P2P) distribution once represented one of the dominant mechanisms for online piracy \cite{mckenzie202025, POUWELSE2008701}, its decline followed increased enforcement and the takedown of major torrent ecosystems such as Pirate Bay \cite{piratebay2014disruption}. increases the perceived risk of detection and legal consequences for users. Consequently, the torrent-based links we observed primarily involved large, bundled content, such as complete television seasons or movie collections, that may exceed the practical storage or transfer limits imposed by Telegram or cloud-hosting providers.

\subsubsection{Bots} 
\label{bots}
Our coders identified and manually analyzed 52 unique bots, referenced 193 times across 29 channels. Similar to the intermediary channels discussed in Section~\ref{multi-channel-qualitative}, these bots either distributed the content directly or redirected users to additional channels and resources. However, unlike channels, bots operate through an interaction-driven model. Rather than exposing content immediately, bots require users to actively initiate a conversation, typically through the \texttt{/start} command, and navigate through menus (or provide subsequent commands) before receiving files or download links.

This interaction-based design has important implications for detection. Many automated abuse detection systems are optimized for static content analysis \cite{mishra2020tacklingonlineabusesurvey, Lewkowicz2025WhenBotsGetSmarter} and may fail to capture malicious behavior without actively interacting with the bot itself. As a result, operators can effectively obscure pirated content behind conversational workflows that are more difficult to analyze at scale.

We further identified 14 bots referenced across multiple channels, with one bot appearing in 6 distinct channels. This suggests the presence of shared distribution infrastructure, where a single bot serves content to users arriving from multiple entry points. Such decoupling of content delivery from channel visibility improves ecosystem resilience: even if a channel is removed, operators can rapidly deploy replacement channels that reconnect users to the same underlying bot infrastructure. Since the mechanisms through which these bots distribute content vary considerably, we dedicate the following section to analyzing their more granular behavioral characteristics.
\subsection{Characterization of Bots}
\label{bots-qualitative}
Our coders manually interacted with 52 bots and categorized them into the following categories:
\newline
\textbf{Content Delivery Bots:} We identified 21 bots that provided direct access to pirated media, either by allowing users to download content within the bot itself or by redirecting them to other channels or external websites. We also noticed that 8 of these bots removed their response (with the download links) within a few minutes after delivering them, and also proactively disabled the forwarding of the response, thus drastically limiting the reach and persistence of the abuse evidence, making it harder for content moderators to trace them later on. 
\newline
\textbf{Dynamic Retrieval Bots:} We observed 13 bots that provided interactive, query-based access to content rather than directly exposing download links. These bots allowed users to search for content by navigating structured menus (similar to directory-style channels) or by entering a search query. For a given query, bots often returned multiple results, including different resolutions, file sizes, or versions of the same content. Figure~\ref{fig:dynamic_retrieval_bot} in the Appendix illustrates one such example, where upon receiving a title name from the user (One Piece, a Japanese anime), the bot returns a list of available episodes across multiple pages. Upon clicking one of the options, the user receives a direct download link for the respective content. Several of them also provide detailed instructions guiding users on how to interact with the system effectively.
\newline
\textbf{Channel Promotion Bots:} We identified 9 bots that required users to join one or more Telegram channels before receiving access to content. After joining the requested channels, the bots provided download links, similar to content delivery bots. Many of the promoted channels were themselves dedicated to video piracy, suggesting that these bots help grow interconnected piracy communities by continuously funneling subscribers across related channels. This is particularly effective since users are often more attracted to communities with high adoption \cite{Fu_Teo_Seng_2012, Katona2011DiffusionSocialNetwork}.
Some promoted channels were explicitly labeled as “backup” channels, allowing operators to preserve subscriber access if the primary bot or channel was removed, while also enabling users to discover alternative channels offering similar content. 

\begin{figure}[ht]
\centering
  \includegraphics[width=0.6\columnwidth]{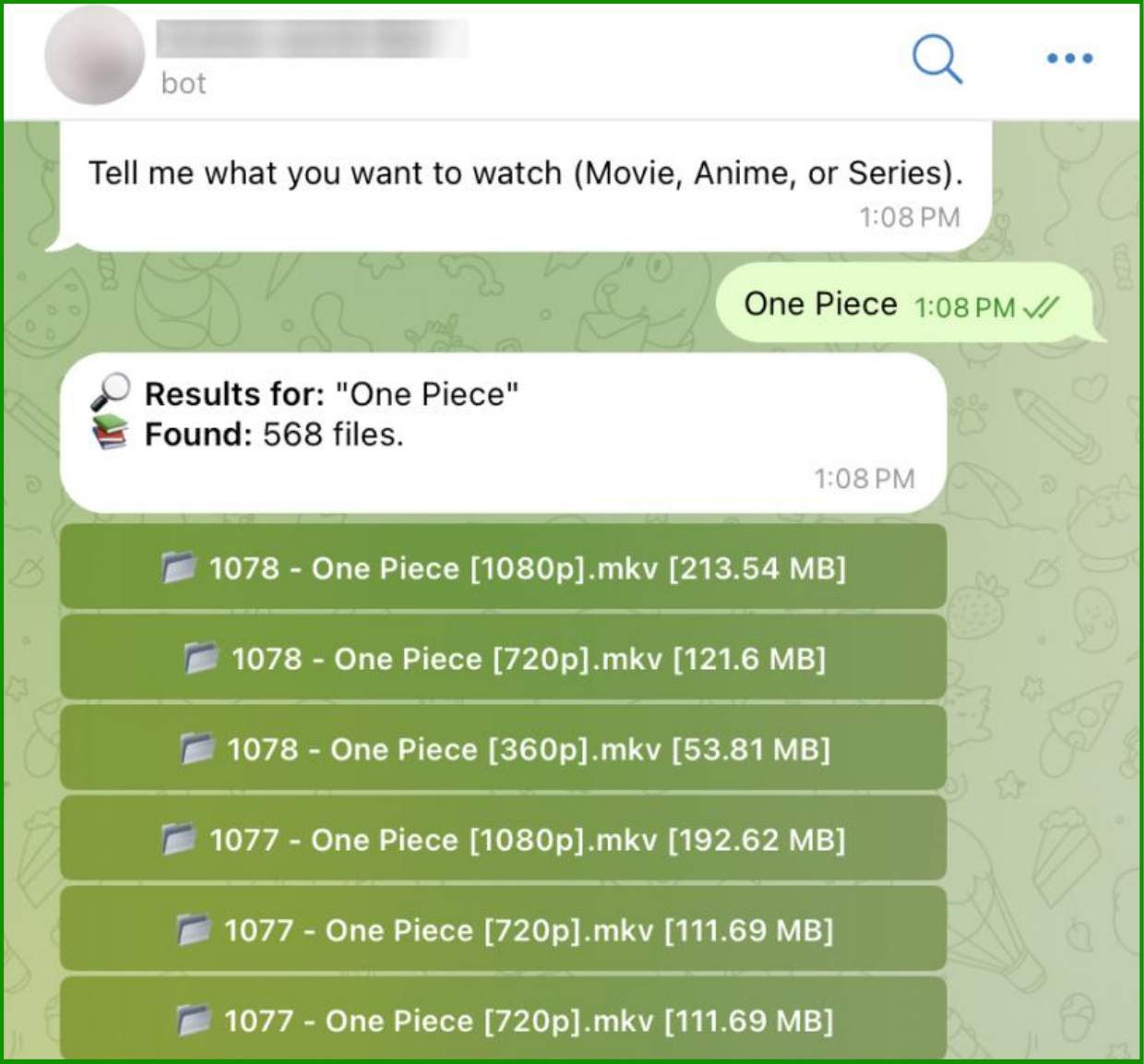}
 \caption{An example of a Dynamic Content retrieval bot}
  \label{fig:dynamic_retrieval_bot}
\end{figure}
\textbf{Content Ingestion Bots:} We discovered 7 bots that allowed users to upload pirated content. These bots enforced structured submission workflows, including restrictions on file types, formatting requirements, and moderation mechanisms such as blacklisting users who repeatedly uploaded invalid or low-quality content. Such controls suggest the presence of quality assurance and verification processes governing submitted material.
Figure~\ref{fig:promotion_and_ingestion_bots}B in the Appendix illustrates one such example.
This suggests that subscribers within these ecosystems act not only as passive consumers, but also as active contributors to content distribution.
Lastly, we identified 2 bots specifically designed to process payment transactions for resources sold by channel operators. We discuss these monetization-oriented bots further in Section~\ref{monetization}.

\begin{figure}[ht]
\centering
  \includegraphics[width=0.6\columnwidth]{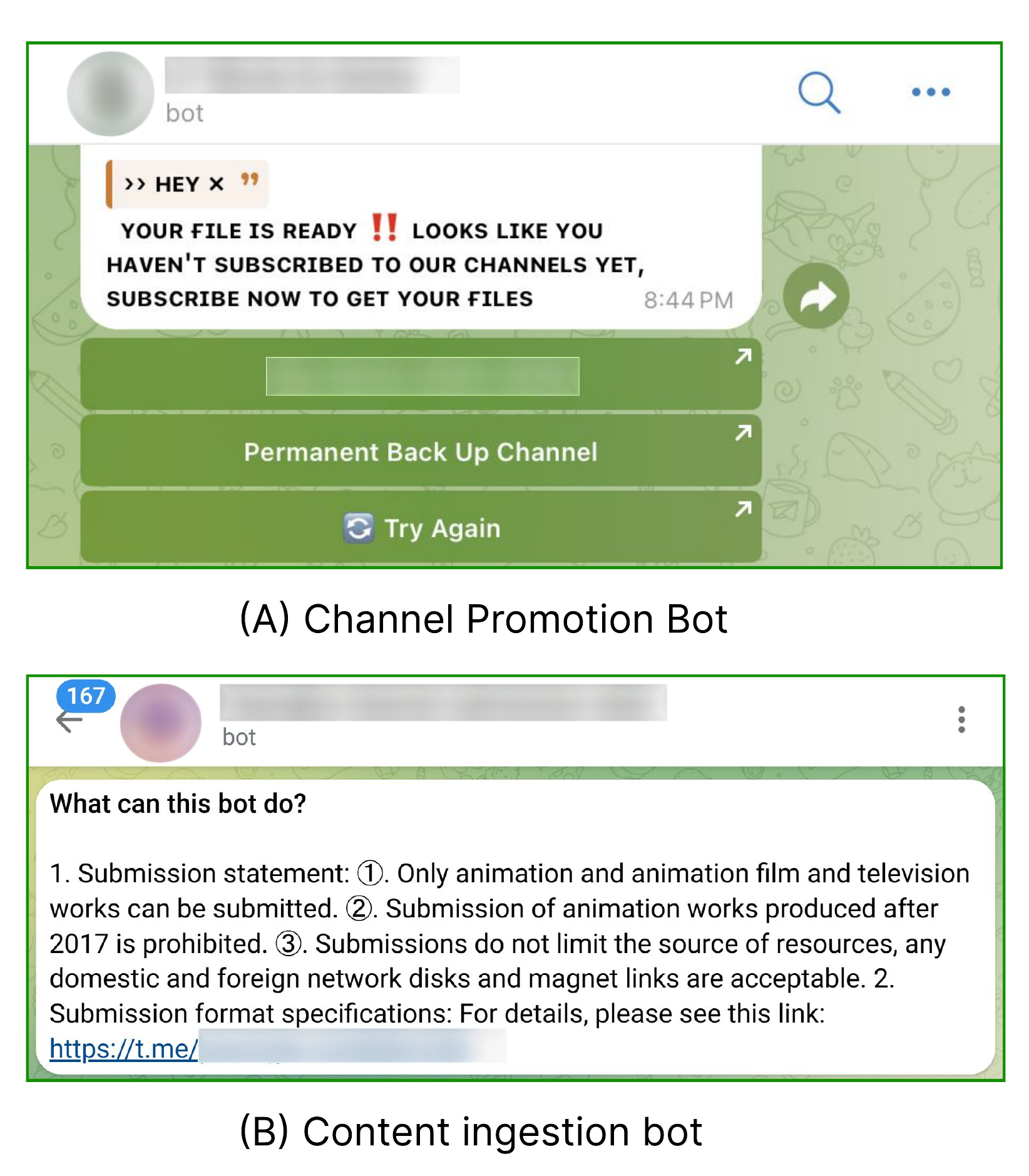}
 \caption{(Top) A channel promotion bot which provides the content only if it's list of channels is followed. (Bottom) A Content ingestion bot which asks users to upload content. }
  \label{fig:promotion_and_ingestion_bots}
\end{figure}

\subsection{Facilitating Illegitimate Access to Legitimate Content} 
We also identified channels that frequently shared resources designed to illegally access content hosted on legitimate streaming platforms. Specifically, we found 18 channels distributing compromised credentials, including usernames, passwords, and session cookies for services such as Netflix, Hulu, Disney+, and Crunchyroll. This aligns with prior work \cite{stayinalive} documenting the distribution of stolen bank accounts and credit card information on Telegram \cite{stayinalive}. \jaishnoor{double check paper content to verify it does mention that}

We additionally observed 11 channels distributing “modded” versions of streaming applications, i.e., modified clients designed to bypass subscription checks and unlock premium content without payment. Notably, all such applications were distributed as Android APK files. Given Telegram’s strong mobile-oriented usage pattern \cite{DigitalWebSolutions2024TelegramStats} and Android’s dominance in the global smartphone market \cite{Kumar2026iPhoneAndroidUsers}, this likely reflects operators tailoring distribution toward the majority of their user base. Beyond compromised accounts and modified applications, we identified 23 channels distributing VPN tools, proxy lists, and IP endpoints intended to bypass geo-restrictions, remove advertisements, or access region-locked content catalogs unavailable in the user’s country. Notably, most posts sharing these resources were accompanied by detailed tutorials, including text guides, annotated screenshots, and instructional videos. This suggests that operators actively lower the technical barrier to entry, enabling even non-expert users to use these tools. 
\subsection{Monetization}
\label{monetization}
Piracy has traditionally been viewed as a “free access” ecosystem~\cite{WhyPay}, where content is redistributed without any payment, with operators only asking for occasional voluntary donations to sustain their operations. However, we observed monetization in two of our seed channels through mechanisms that are directly embedded within the distribution pipeline. In one channel, users were required to purchase credits to download content from the external link (Terrashare). As noted in Section~\ref{external_links_qualitative}, cloud providers such as Terrashare operate on a model where users need to expend credits to download the content. We also found that this channel incentivized users to contribute uploads by rewarding them with credits transferred to their cloud accounts, aligning with our observations in Section~\ref{bots-qualitative}, where bots actively solicited user uploads.\josef{How are these credits tracked? Is it on the user's Terrashare account or somehow within Telegram? As a reader not familiar with monetization on Telegram, I would have this question.} In another channel, monetization was tied to content quality. While content up to 1080p was made freely available, users were required to pay to access higher-resolution versions such as 4K.

In both cases, these channels utilized a bot to facilitate payment. As discussed in Section~\ref{bots}, bots provide a key advantage by dynamically revealing payment instructions only upon user interaction, making these workflows largely invisible to static abuse detection systems. Furthermore, these bots use Telegram’s embedded payment API, which enables them to process transactions through mobile and cryptocurrency wallets such as Google Pay, Apple Pay, MetaMask, and Coinbase. This abstraction allows users to complete payments without directly sharing sensitive financial information with channel operators, while simultaneously removing the need for operators to deploy and maintain external payment infrastructure. By acting as an intermediary, Telegram further complicates transaction tracing and attribution for abuse detectors.

%% file: sections/taxonomy.tex
\section{Taxonomy}  
\label{taxonomy}

\begin{figure*}[t]
\centering
  \includegraphics[width=0.9\textwidth]{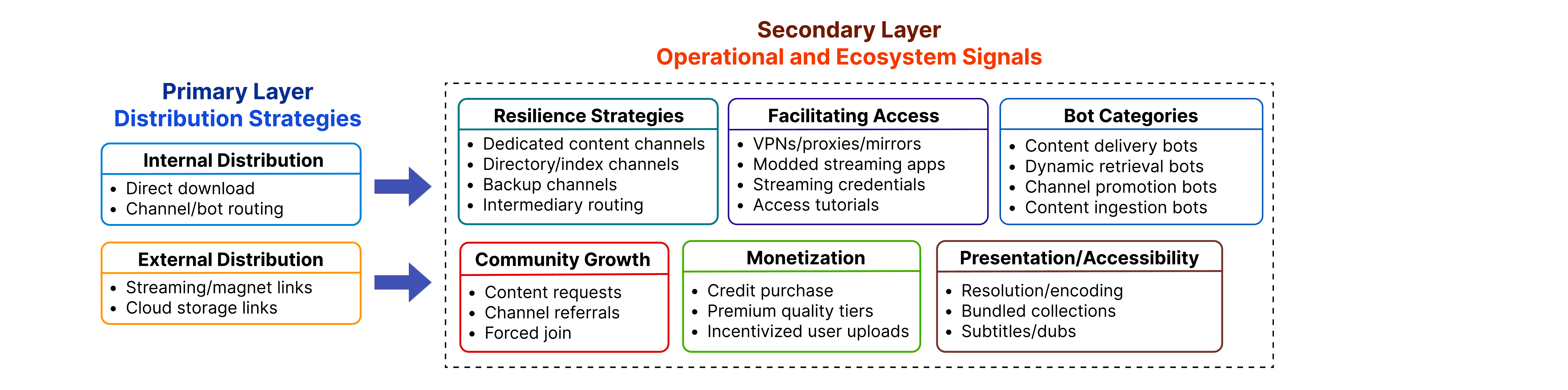}
\caption{Taxonomy of piracy-related posts on Telegram derived from our insights in Section~\ref{qualitative_analysis}.}
  \label{fig:taxonomy}
\end{figure*}

Our qualitative analysis in the previous section revealed that video piracy on Telegram is not a homogeneous process, with channels utilizing a diverse set of behaviors to distribute, organize, monetize, and sustain access to pirated content. Modeling such activity using only a binary “piracy” label can therefore lead to a substantial loss of informational granularity, thus obscuring the structural characteristics of the broader ecosystem. To address this, we develop a structured taxonomy that captures the fine-grained attributes of piracy-related posts. The taxonomy is grounded in our qualitative analysis of piracy seed channels, with the two independent coders identifying recurring behavioral and operational patterns and organizing them into a coherent hierarchy of labels. The coders achieved strong inter-rater agreement (Cohen’s Kappa = 0.82), with subsequent disagreements resolved through discussion. Figure~\ref{fig:taxonomy} illustrates the resulting taxonomy.

We designate \textbf{Distribution Strategies} as the \emph{Primary Layer} of our taxonomy, since the core objective of piracy channels is ultimately the distribution of copyrighted content. This layer captures the principal delivery pathways used by channels and is divided into two broad categories: \textbf{Internal Distribution}, where content is delivered within Telegram itself either through direct file uploads or by routing through one or more channels or bots, and \textbf{External Distribution}, where channels redirect users to cloud-storage services, streaming websites, and torrent or magnet links. 

Complementing this primary layer, we introduce a \textbf{Secondary Layer} consisting of \textbf{Operational and Ecosystem Signals}, which captures the broader mechanisms through which piracy communities organize, sustain growth, evade disruption, and facilitate user access.
First, we identify \textbf{Resilience Strategies}, which capture how channels can evade takedown efforts. These include the use of dedicated content channels, directory or index channels, backup channels, and intermediary routing.
Second, the layer captures \textbf{Facilitating Access} behaviors, which enable users to illegally gain access to legitimately hosted content by sharing VPNs, proxies, mirrors, modded streaming applications, compromised streaming credentials, and detailed access tutorials or setup guides.
Third, considering the important role bots play in automating and scaling piracy operations on Telegram, we also add a "Bot Categories" section which captures all four bot variants we observed in our seed dataset: content delivery bots, dynamic retrieval bots, channel promotion bots, and content ingestion bots.
The taxonomy also captures ecosystem-level behaviors related to \textbf{Community Growth} including content requests, where operators request subscribers to upload new titles, channel referrals that ask subscribers to join channels, and forced joins where bots only provide access to content if the user follows the listed channels. The taxonomy also captures \textbf{Monetization} signals we observed in Section~\ref{monetization}, which include operators asking subscribers to purchase credits (to download content from cloud-sharing services), providing premium quality content for a fee, and also incentivizing subscribers for uploading new content. Finally, the taxonomy captures \textbf{Presentation and Accessibility} signals that describe how pirated content is packaged and presented to users as observed in Section~\ref{content_metadata}. These include resolution and encoding information, bundled collections, and the presence of subtitles or dubbed variants.

Building this taxonomy has two direct implications. \textit{First,} it provides a structured foundation for assessing the extent to which the characteristics identified in our qualitative analysis manifest at scale, enabling systematic measurement of their prevalence within the larger quantitative dataset analyzed in Section~\ref{quantitative_analysis}. \textit{Second,} the taxonomy enables the development of detection systems capable of providing stakeholders, including content rights holders, security vendors, and Telegram itself, with rich contextual information about piracy-related activity to support timely and effective intervention, an effort we operationalize in Section~\ref{real_word_impact}.

%% file: sections/quantitative_analysis.tex
\section{Large-Scale Measurement of the Telegram Piracy Ecosystem}
\label{quantitative_analysis}
To complement our qualitative findings, we next conduct a large-scale quantitative analysis of the Telegram piracy ecosystem. The goal of this section is twofold: first, to evaluate whether the behavioral patterns identified during the qualitative analysis generalize at scale, thereby validating the practical utility of our taxonomy; and second, to analyze ecosystem characteristics that were difficult to observe through the smaller qualitative seed set alone, such as large-scale network connectivity, cross-channel coordination, and transnational redistribution patterns.

\subsection{Data Collection}
\label{data_collection}

% \begin{table*}[ht]
% \centering
% \begin{tabular}{l|ccc|ccc|ccc}
% \hline
% \textbf{Channel category} 
% & \multicolumn{3}{c|}{\textbf{Posts per channel}} 
% & \multicolumn{3}{c|}{\textbf{Subscribers}} 
% & \multicolumn{3}{c}{\textbf{Total views per channel}} \\
% \cline{2-10}
% & Min/Max & Median & Mean 
% & Min/Max & Median & Mean 
% & Min/Max & Median & Mean \\
% \hline

% Initial hydrated channels (489) 
% & 1/500 & 284 & 248 
% & -- & -- & -- 
% & 0/43,361,599 & 496,170 & 2,315,217 \\

% 1-hop channels (233) 
% & 1/497 & 207 & 219 
% & -- & -- & -- 
% & 13/875,919,997 & 327,500 & 5,775,149 \\

% 2-hop channels (261) 
% & 1/438 & 271 & 228 
% & -- & -- & -- 
% & 37/1,493,059,999 & 140,448 & 7,158,879 \\

% Merged dataset (983) 
% & 1/500 & 263 & 236 
% & -- & -- & -- 
% & 0/1,493,059,999 & 309,250 & 4,421,382 \\

% \hline
% \end{tabular}
% \caption{Descriptive statistics across datasets (Initial, 1-hop, 2-hop, and merged). All statistics are computed per channel.}
% \label{tab:descriptive_statistics}
% \end{table*}
\subsubsection{Discovering Channels}
\label{discovering_channels}
Unlike platforms such as Facebook and X~\cite{fbgraphsearch}, Telegram requires explicit channel names to find them, making discovery through fuzzy search approaches challenging. To overcome this limitation, we utilized SearXNG~\cite{searxng}, an open-source metasearch engine that aggregates results from multiple providers, including Google, Bing, and Yahoo, to identify Telegram channels that reference movies and TV shows.

To construct the search queries, we used the TMDb dataset~\cite{tmdb}, a large-scale repository of movie and television metadata. We filtered this dataset to include titles released from the year 1980 onward, resulting in 247,507 movies and 141,229 TV shows spanning 152 countries. To reduce noise, we excluded titles containing common English phrases (e.g., generic or ambiguous terms such as ``Home,'' ``Love,'' or ``You'') that are likely to yield non-specific or irrelevant results. We then exhaustively queried each title in January 2026 using the pattern \fbox{\footnotesize\texttt{site:t.me "<content\_name>"\&time\_range=week}}, which retrieves Telegram channels referencing the content within the past 7 days. After deduplicating results that mapped to the same channel, this process yielded the links for 15,198 unique Telegram channels. It is worth noting that although these channels were recently indexed by search engines, this does not imply that they are newly created; rather, they may have only been discovered or surfaced recently.
\subsubsection{Identifying Video Piracy Channels}
\label{identifying_piracy_channels}
\begin{figure}[ht]
\centering
  \includegraphics[width=0.9\columnwidth]{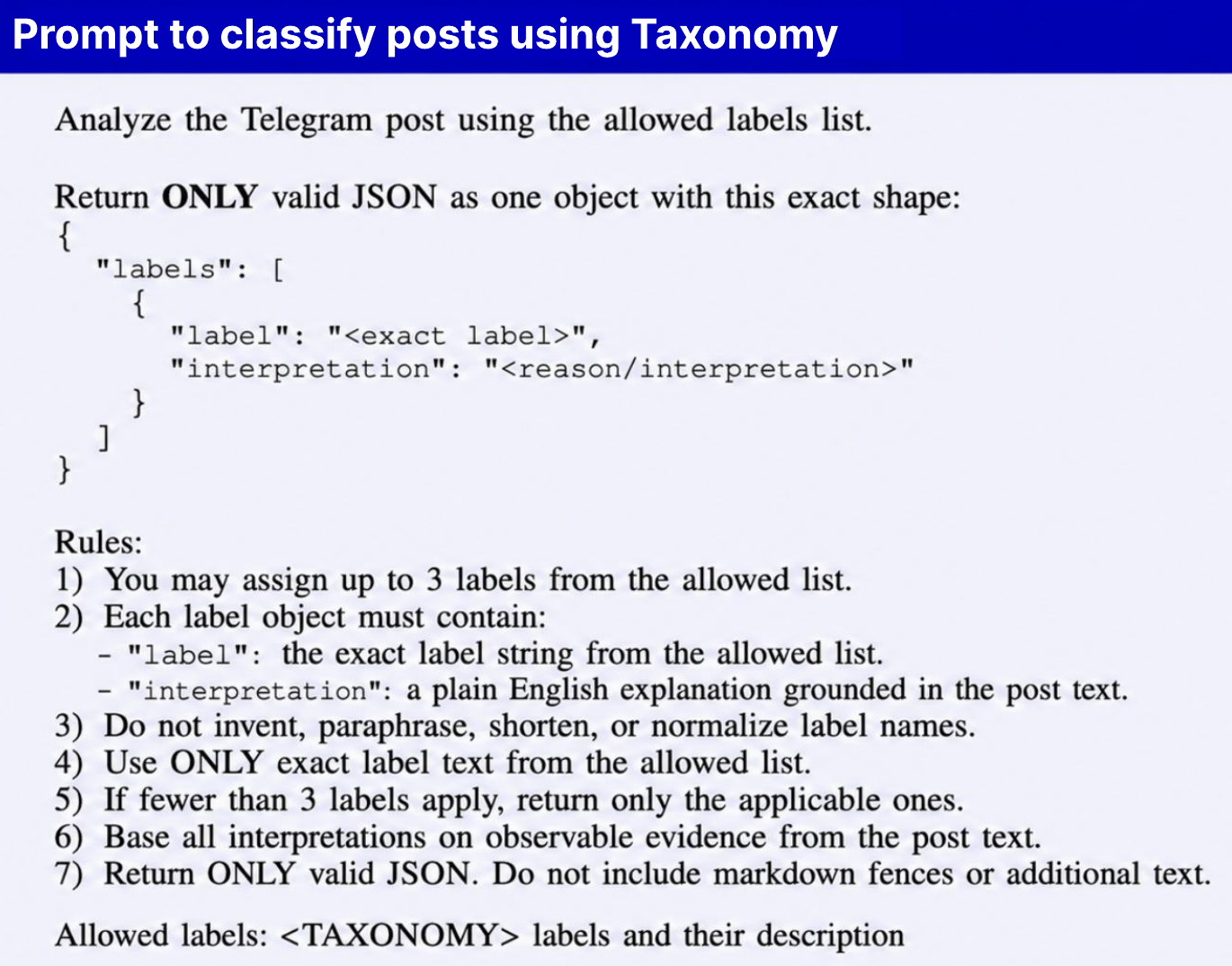}
  \caption{Prompt used to classify posts using Taxonomy.}
  \label{fig:taxonomy_prompt}
\end{figure}

The presence of a content name alone does not imply that the channel distributes pirated content. Many channels reference movies and TV shows for legitimate purposes, such as news updates, trailers, reviews, or fan discussions.  Thus, to identify which of the 15,198 collected channels were engaged in video piracy, we first sampled 10 posts from each channel. We then applied Gemma3:27B, a locally deployed Large Language Model (LLM), to label these posts based on the taxonomy defined in Section~\ref{taxonomy}. Recent work has shown that LLMs can effectively support large-scale annotation tasks when provided with well-defined instructions.~\josef{cite?} For each post, the model first determines whether it constitutes piracy according to the behavioral definitions in our taxonomy. If classified as piracy, the model assigns up to three taxonomy labels and provides a justification for each assignment. If the post is deemed non-piracy, it is labeled as benign, along with an explanation supporting this decision. The full prompt used for labeling is provided in Fig.~\ref{fig:taxonomy_prompt}.

To evaluate labeling accuracy, two human coders independently reviewed a random sample of 1,000 posts, assessing both the piracy-versus-benign classification and the correctness of the assigned taxonomy labels. The coders had a Cohen's Kappa inter-rater agreement of 0.82, signifying almost perfect agreement, and after solving disagreements, found that the model was 99.2\% accurate at identifying whether the post constituted piracy, and 97\% accurate at correctly identifying taxonomy labels. They also identified only 4 benign samples which were marked as piracy. Overall, we identified 489 channels where one or more posts were distributing pirated content. 
\subsubsection{Hydrating channels}
\label{hydrating_channels}
Consistent with Section~\ref{qualitative_analysis} and to preserve recency in the observed content, we collected up to 500 recent posts per channel from the 489 identified channels, spanning December 2023 to January 2026.
Given that these channels frequently distribute content via intermediary channels and bots, we expanded our collection by recursively hydrating all linked entities up to a depth of two, including channels and bots referenced by the original channels. This process resulted in a total of 508 channels, of which 494 were active, and 1,151 bots, of which 1,066 were active. Notably, this contrasts with our qualitative seed set of 74 channels (Section~\ref{qualitative_analysis}), where a majority of linked channels and bots were already inactive at the time of analysis. Thus, a much higher proportion of active links in this dataset enables a more comprehensive and accurate analysis of how channels interconnect within the ecosystem, which we explore later in Section~\ref{network_analysis}.
All posts in these channels were also subsequently labeled by Gemma3:27B using the taxonomy. Similar to our seed dataset in Section~\ref{qualitative_analysis}, the text in the posts across the channels comprised of several languages. To ease analysis, we utilized TranslateGemma:12B, a variant of Gemma3 highly regarded as a translation model, to translate all posts in the dataset. 
Overall, we collected 983 channels (489 original, 233 at depth 1, and 261 at depth 2). 
% and analyze their impact on user interaction patterns in Section~\ref{user_activity}\josef{Do we have a section on user activity? Some of it (views, forwards) is already mentioned in this section.}.

\subsection{Content Distribution Landscape}
\label{content_distribution_landscape}
To understand the scale at which pirated titles are shared within these channels, we first identify the specific content that was being distributed, and then identify production companies and countries of origin. 
Similar to~\ref{data_collection}, we searched for entries from the TMDb across all the channel posts using Gemma3:27B. We utilize an LLM instead of a simple string match to allow for fuzzy search. As evident in our qualitative analysis in Section~\ref{qualitative_analysis}, not all posts directly refer to the content name; instead, they redirect to other channels or share circumvention resources. Also, the same content name may be released by different distributors across countries (for example, ``The Office" UK  and US were produced by different studios (BBC Two and NBC, respectively).

\begin{figure}[ht]
    \centering
    \includegraphics[width=0.8\columnwidth]{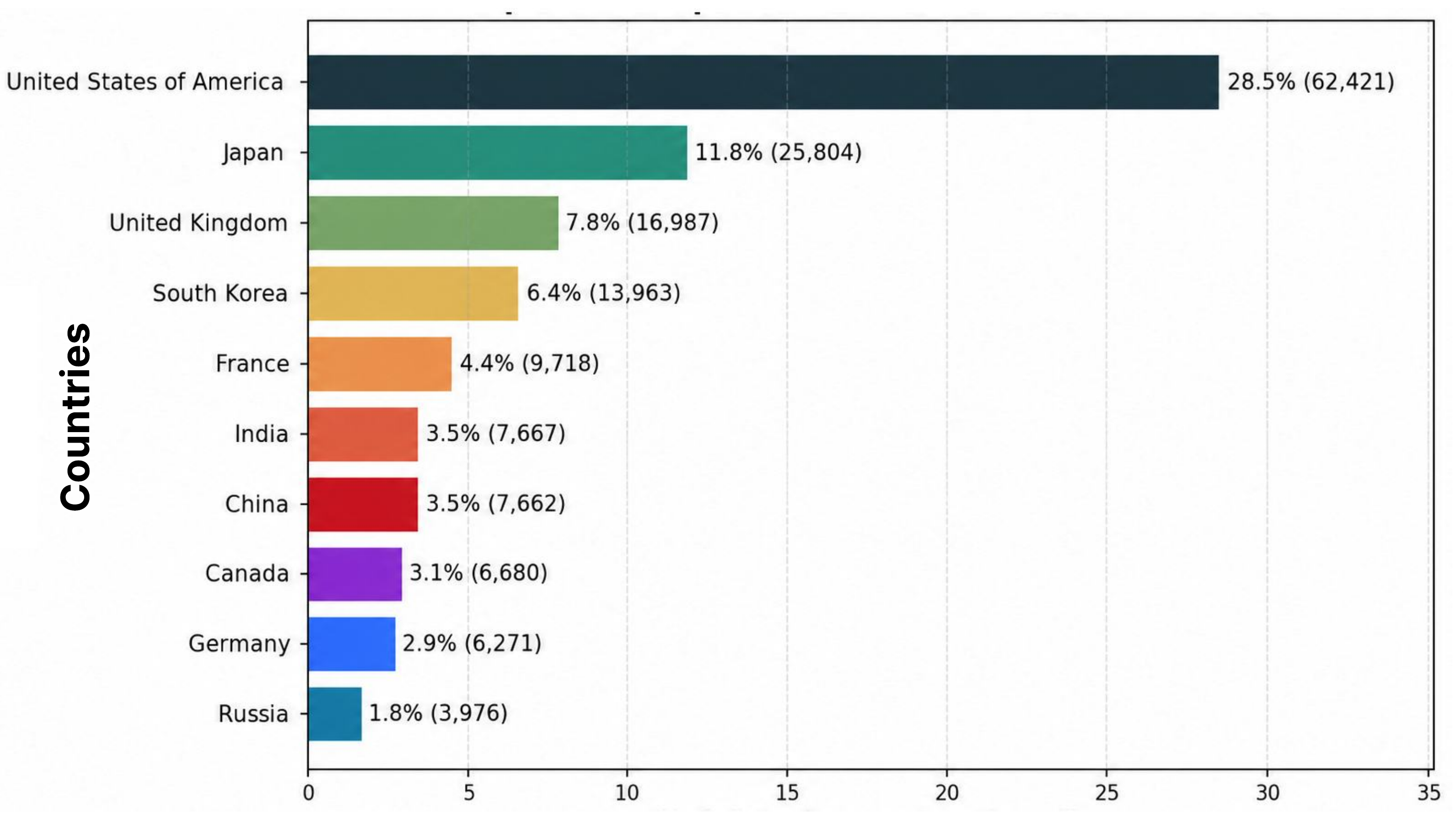}
    \sayak{Have numbers, and then the percentage in brackets}
    \caption{Top 10 production countries whose content was pirated in the Telegram channels}
    \label{fig:production-countries}
\end{figure}

\begin{figure}[ht]
    \centering
    \includegraphics[width=0.8\linewidth]{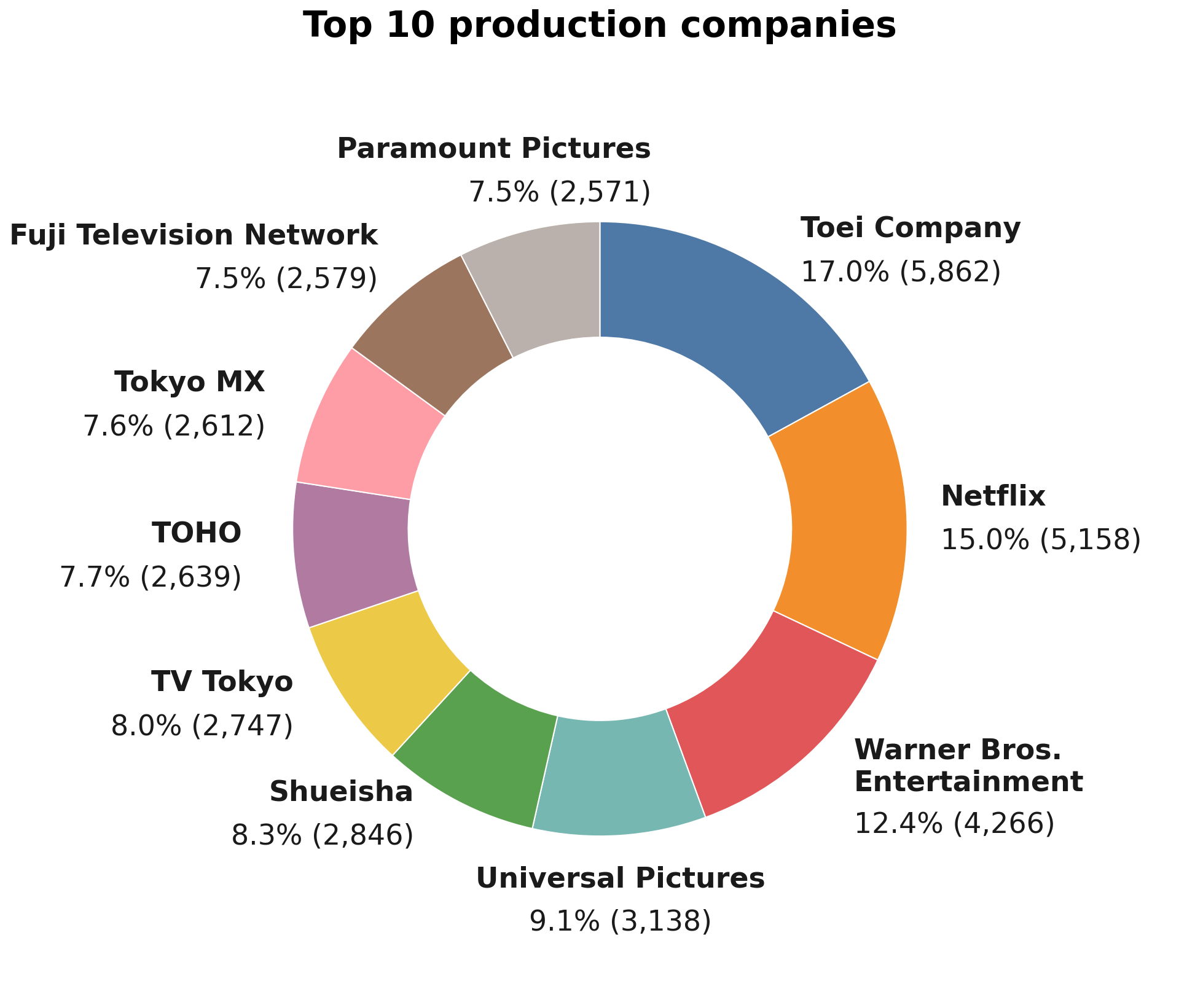}
    \caption{Distribution of the top 10 production companies whose content was pirated.}
    \label{fig:production-companies}
\end{figure}

Using this approach, we identified 240,943 content matches across the posts in these channels, which correspond to 19,033 unique titles (14,632 movies and 4,401 TV shows) produced by 3,941 companies in 118 unique countries. Two coders again manually validated 1,000 such randomly selected labels and identified 994 of them to be correct, ensuring minimal noise across the content identification process. Figure~\ref{fig:production-companies} illustrates the top 10 production companies whose content was pirated in these Telegram channels, with content produced by the Toei Company (17\%, or 5,862 unique titles) being the most pirated, followed closely by Netflix (15\%) and Warner Bros. Discovery (12.4\% or 4,266 unique titles). We also find that 28.5\% of all content pirated was produced in the United States, whereas 11.8\% was in made Japan, followed by United Kingdom (7.8\%) and South Korea (6.4\%). Figure~\ref{fig:production-countries} illustrates the top 10 countries that produced the titles that were pirated. It is worth noting that the majority of the content originates from countries that have the strictest piracy enforcement (US, Japan, South Korea)~\cite{AFP2016JapanPiracy, 
JustiaPiracyEntertainment}.

Next, we estimated the countries from which the piracy posts/channels originated. We again used TranslateGemma:12B to identify the language for posts in each channel. Table~\ref{tab:language_country_confusion} illustrates the confusion matrix between the inferred posting language of the channels and the production country of the pirated content. We find that a substantial proportion of content produced in the United States is redistributed through Persian-language channels (32,960 posts), significantly exceeding redistribution through English-language channels themselves (10,768 posts). A similar trend is also observed for content originating from the United Kingdom and France, suggesting that Western-produced media is heavily redistributed within non-Western language ecosystems.
We also observe strong regional specialization in the distribution of East Asian content. Japanese-produced titles are highly prevalent within both Persian-language (8,674) and Chinese-language (4,079) channels, while South Korean content is most frequently distributed through Burmese-language channels (5,189), followed by Persian-language channels (4,382). 

Interestingly, English-language channels do not dominate the redistribution of content produced in English-speaking countries. Instead, much of this content appears to be redistributed through channels associated with regions where legal access may be restricted \cite{LuRajaviDinner2021OTTpiracy, Lynch2025GeoRestrictedStreaming, Boulifi2026PiracyProblemStreaming}, subscription costs may be comparatively expensive \cite{Ibeh2025NetflixNigeriaAffordability, Broderick2025StreamingPiracyReturn}, or enforcement mechanisms may be weaker. Similarly, Chinese-language, Persian-language, and Burmese-language \josef{Shouldn't Burmese get a mention?} channels consistently appear among the dominant redistributors across nearly all production countries shown in Table~\ref{tab:language_country_confusion}, indicating that channels originating from these regions play a central role in the global dissemination of pirated media on Telegram.

Overall, these findings suggest that Telegram piracy ecosystems are highly transnational in nature. Content produced in one country is frequently redistributed through channels operating in entirely different linguistic and regional contexts, enabling piracy operators to bypass geographic licensing restrictions and extend the reach of copyrighted media to audiences far beyond its intended distribution markets.
\begin{table*}[htbp]
\centering
\caption{Cross-regional redistribution patterns between channel posting language and the production country of pirated content. Rows represent the country in which the content was produced, while columns represent the inferred primary language of the Telegram channels distributing that content.}
\label{tab:language_country_confusion}
\resizebox{0.9\textwidth}{!}{%
\begin{tabular}{lrrrrrrrrrr}
\toprule
Production Country & Persian & English & Burmese & Chinese & Arabic & Spanish & Russian & Portuguese & French & Hindi \\
\midrule
United States of America & 32960 & 10768 & 5251 & 4732 & 2597 & 2013 & 1761 & 1543 & 1118 & 384 \\
Japan & 8674 & 8332 & 1968 & 4079 & 1124 & 1131 & 315 & 778 & 198 & 691 \\
United Kingdom & 9625 & 3008 & 1363 & 1476 & 633 & 411 & 478 & 336 & 253 & 110 \\
South Korea & 4382 & 2806 & 5189 & 1616 & 346 & 358 & 169 & 178 & 90 & 376 \\
France & 5379 & 1503 & 1015 & 834 & 331 & 230 & 218 & 200 & 251 & 51 \\
% China & 3015 & 1644 & 1829 & 2502 & 197 & 123 & 156 & 93 & 44 & 208 \\
% India & 3563 & 1586 & 890 & 862 & 236 & 133 & 119 & 108 & 51 & 416 \\
% Canada & 3817 & 1313 & 745 & 575 & 254 & 172 & 218 & 188 & 146 & 43 \\
% Germany & 3561 & 1005 & 743 & 528 & 254 & 210 & 188 & 128 & 113 & 31 \\
% Russia & 2323 & 786 & 604 & 685 & 177 & 89 & 120 & 77 & 30 & 18 \\
\bottomrule
\end{tabular}}
\end{table*}

\subsection{Financial damage}
\label{financial_damage}
One of the primary consequences of video piracy is the financial loss incurred by content rights holders through reduced digital and physical media sales, as well as streaming subscriptions. These losses also downstream to independent creators, distributors, and industry professionals whose livelihoods depend on legitimate consumption of the content they help create.

In Section~\ref{content_distribution_landscape}, we identified 19,033 unique titles distributed across the observed channels. In this section, we estimate the approximate financial impact associated with the distribution and consumption of this pirated content. To approximate this impact, we adopt a conservative methodology, inspired by Roy et al.~\cite{roy2025darkgram} when estimating damage caused by software piracy, of assuming that 1\% of post views correspond to actual content consumption. That is, for a post with 100 views, we assume one user accessed the content and represents a potential loss in revenue. This assumption reflects realistic user behavior, where individuals typically browse multiple posts but consume only a small subset of the available content.

Estimating financial loss for video content is inherently complex due to the diversity of legal access models. Unlike software piracy, which typically follows a fixed pricing structure, video content can be accessed through subscriptions, rentals, purchases, or bundled services. To address this, for each post referencing a matched title, we utilized the Claude-Haiku model with the web\_search\_20260209 extension to retrieve real-time availability and pricing information. We opted for a retrieval-enabled model instead of a local model (e.g., Gemma3:27B) because content availability and pricing vary dynamically across regions and platforms, requiring up-to-date external knowledge.
For each title, we first inferred the likely country of distribution based on the language of the post, as described in Section~\ref{content_distribution_landscape}. We then queried whether the title was legally available on streaming platforms within that region. When multiple pirated titles were available on the same streaming service in the same region, we grouped them together and estimated the loss using the lowest subscription cost for that service, rather than assigning a separate cost per title, since a user could have accessed all those titles through one subscription. We then applied the 1\% view-to-consumption assumption at this grouped, subscription level.
For titles that were not available on any identified streaming platform in the corresponding region, we estimated the loss using the lowest available legal access cost, prioritizing rental prices and falling back to the lowest physical media price (e.g., DVD) when rental options were unavailable. All monetary values were normalized to USD using exchange rates as of January 2026.
To validate the accuracy of the pricing and availability extraction, we manually evaluated 1,000 sampled instances, finding that 971 were correctly identified, indicating high reliability with minimal noise. Overall, we estimate a total financial loss of \textbf{\$17.49B} attributable to content shared within our dataset. Table~\ref{tab:financial_loss_countries} presents a breakdown of the top five affected countries, separated by movies and television content.

\begin{table}[ht]
\centering
\small
\caption{Estimated lower-bound financial losses associated with pirated content distributed through Telegram channels, grouped by the production country of the affected titles.}
\label{tab:financial_loss_countries}
\resizebox{0.75\columnwidth}{!}{%
\begin{tabular}{lrr}
\toprule
\textbf{Country} & \textbf{Movies (USD)} & \textbf{Television (USD)} \\
\midrule
United States & \$4.55B & \$3.62B \\
Japan & \$2.14B & \$1.58B \\
South Korea & \$1.08B & \$1.41B \\
United Kingdom & \$0.84B & \$0.79B \\
India & \$0.38B & \$0.29B \\
\midrule
Top 5 Total & \$8.99B & \$7.69B \\
\bottomrule
\end{tabular}}
\end{table}

Even under these conservative assumptions, the results indicate substantial global revenue losses associated with video piracy on Telegram. We find that content produced in the United States experiences the highest estimated financial impact, accounting for approximately \$8.17B in potential displaced revenue, followed by Japan at \$3.72B. As previously observed in Table~\ref{tab:language_country_confusion}, a large proportion of content originating from these countries is redistributed and consumed through channels associated with countries such as Iran, China, Myanmar, and Russia, regions that have historically faced weaker copyright enforcement, limited anti-piracy oversight, or inconsistent regulation surrounding digital content distribution.

\subsection{Taxonomy-Guided Ecosystem Characterization}
\label{ecosystem_characterization}

Using the taxonomy labels assigned to each post, we next quantify the behavioral and operational patterns identified during the qualitative analysis across the substantially larger dataset. The goal of this analysis is to examine whether the ecosystem characteristics observed qualitatively persist at scale throughout the broader Telegram piracy ecosystem. To achieve this, we quantify the prevalence of posts associated with each taxonomy category. Table~\ref{tab:taxonomy_grouped} illustrates the distribution of these taxonomy categories across the dataset, including the total number of posts, the number of unique channels in which each behavior appeared, and the average and median frequency of occurrence per channel.
\\
\textbf{Internal Distribution:} Directly downloadable files hosted within Telegram itself appeared in 6,589 posts from 440 unique channels, whereas channel/bot-routing behaviors appeared in 3,843 posts from 337 unique channels, reinforcing our qualitative observation that Telegram piracy ecosystems heavily rely on Telegram-native infrastructure and interconnected routing pipelines for large-scale content delivery. We further explore the structural characteristics of these interconnected distribution pipelines in Section~\ref{network_analysis}.
\\
\textbf{External Distribution:} We identified 3,982 posts from 154 unique channels associated with 33 unique cloud-storage providers. Similar to our qualitative observations, the majority of these links pointed toward less prominent providers such as \texttt{cloud.189.cn} (994 posts), \texttt{alipan.com} (972 posts), and \texttt{pan.quark.cn} (760 posts), whereas mainstream providers such as \texttt{MediaFire} (132 posts), \texttt{iCloud} (27 posts), and \texttt{Google Drive} (25 posts) appeared only rarely. This further supports our earlier finding that operators intentionally favor lower-profile cloud-hosting services, likely due to reduced moderation pressure and lower operational costs. Additionally, we identified 2,043 streaming and magnet-link posts from 195 unique channels shared throughout the dataset. 
\\
\textbf{Resilience Strategies:} We observe strong evidence of resilience-oriented behaviors throughout the ecosystem. Dedicated content channels appeared in 4,575 posts from 466 unique channels, whereas directory/index channels appeared in 6,686 posts from 773 unique channels, backup channels in 571 posts from 232 unique channels, and intermediary-routing behaviors in 1,624 posts from 211 unique channels. Collectively, these findings strongly support our qualitative observation that operators deliberately structure distribution pipelines to remain operational despite moderation efforts and partial takedowns.
\\
\textbf{Facilitating Access:} We also identified widespread behaviors aimed at facilitating unauthorized access to legitimate streaming ecosystems. Access-facilitation and circumvention behaviors appeared in 768 posts from 104 unique channels involving VPNs, proxies, or mirror resources. We also identified 3,279 posts from 399 unique channels advertising modified streaming applications and 585 posts from 143 unique channels sharing compromised streaming credentials. Additionally, instructional tutorials and setup guidance appeared in 1,571 posts from 269 unique channels, suggesting that operators actively lower the technical barriers associated with piracy consumption through walkthroughs and access guidance.
\\
\textbf{Community Growth:} Community-driven participation also generalized strongly at scale. Channel-referral behaviors appeared in 2,973 posts from 273 unique channels, while forced-join mechanisms appeared in 884 posts from 118 unique channels. Similarly, content-request behaviors appeared in 2,299 posts from 608 unique channels. These findings suggest that subscribers do not operate solely as passive consumers, but actively participate in shaping the ecosystem by requesting, sourcing, and redistributing pirated content.
\\
\textbf{Monetization:} Although monetization-oriented behaviors appeared less frequently than distribution-related behaviors, they were still consistently present throughout the dataset. Credit-purchase mechanisms appeared in 9,995 posts from 435 unique channels, premium-quality access tiers in 3,128 posts from 268 unique channels, and incentivized upload behaviors in 1,742 posts from 189 unique channels. These findings suggest that portions of the ecosystem increasingly integrate monetization directly into the content distribution pipeline itself.
\\
\begin{table}[t]
\centering
\caption{Distribution of taxonomy categories assigned to the posts in Section~\ref{quantitative_analysis}.}
\label{tab:taxonomy_grouped}
\resizebox{0.8\columnwidth}{!}{%
\begin{tabular}{@{}lrrr@{}}
\toprule
\textbf{Category} & \textbf{Count} & \textbf{Ch.} & $\mathbf{\bar{x}/\tilde{x}}$ \\
\midrule

\multicolumn{4}{@{}l@{}}{\fcolorbox{blue}{white}{\textbf{Internal Distribution}}} \\
Direct download       & 6,589 & 440 & 14.97 / 5.00 \\
Channel/bot routing   & 3,843 & 337 & 11.40 / 3.00 \\

\addlinespace[2pt]
\multicolumn{4}{@{}l@{}}{\fcolorbox{orange}{white}{\textbf{External Distribution}}} \\
Streaming/magnet links & 2,043 & 195 & 10.48 / 1.31 \\
Cloud storage links    & 3,982 & 154 & 25.86 / 4.50 \\

\addlinespace[2pt]
\multicolumn{4}{@{}l@{}}{\fcolorbox{teal}{white}{\textbf{Resilience Strategies}}} \\
Dedicated content channels & 4,575 & 466 & 9.82 / 4.00 \\
Directory/index channels   & 6,686 & 773 & 8.65 / 2.79 \\
Backup channels            & 571   & 232 & 2.46 / 1.00 \\
Intermediary channels      & 1,624 & 211 & 7.70 / 2.00 \\

\addlinespace[2pt]
\multicolumn{4}{@{}l@{}}{\fcolorbox{violet}{white}{\textbf{Facilitating Access}}} \\
VPNs/proxies/mirrors  & 768   & 104 & 7.38 / 2.00 \\
Modded streaming apps & 3,279 & 399 & 8.22 / 2.80 \\
Streaming credentials & 585   & 143 & 4.09 / 1.00 \\
Access tutorials      & 1,571 & 269 & 5.84 / 2.00 \\

\addlinespace[2pt]
\multicolumn{4}{@{}l@{}}{\fcolorbox{red}{white}{\textbf{Community Growth}}} \\
Content requests & 2,299 & 608 & 3.78 / 1.63 \\
Channel referrals & 2,973 & 273 & 10.89 / 2.00 \\
Forced join       & 884   & 118 & 7.49 / 1.00 \\

\addlinespace[2pt]
\multicolumn{4}{@{}l@{}}{\fcolorbox{green}{white}{\textbf{Monetization}}} \\
Credit purchase            & 9,995 & 435 & 22.98 / 4.00 \\
Premium quality tiers      & 3,128 & 268 & 11.67 / 3.50 \\
Incentivized user uploads  & 1,742 & 189 & 9.22 / 2.50 \\

\addlinespace[2pt]
\multicolumn{4}{@{}l@{}}{\fcolorbox{brown}{white}{\textbf{Presentation/Accessibility}}} \\
Resolution/encoding & 50,454 & 1,114 & 45.29 / 10.30 \\
Bundled collections & 22,113 & 609   & 36.31 / 12.00 \\
Subtitles/dubs      & 18,796 & 458   & 41.04 / 9.00 \\

\bottomrule
\end{tabular}%
}
\end{table}
\newline
\textbf{Presentation and Accessibility:} We found that bundled collections appeared in 22,113 posts from 609 unique channels. Additionally, 50,454 posts from 1,114 unique channels included multiple resolutions or encoding variants, while 18,796 posts from 458 unique channels provided subtitles or dubbed versions of the same content. These findings further reinforce our qualitative observation that operators actively optimize content presentation and accessibility to appeal to broader international audiences.
\newline
Overall, the quantitative findings strongly validate the patterns identified in our qualitative analysis, reaffirming not only the complexity and depth of Telegram's video piracy ecosystem but also the practical utility of our taxonomy in systematically characterizing the behavioral patterns that emerge within these communities at scale.

%%%%%%%%%%%%%%%%%%%%%%%%%%%%%%%%%%%%%%%%%%%%
%%%%%%%%%%%%%%%%%%%%%%%%%%%%%%%%%%%%%%%%%%%%

\subsection{Internal Redirection and Network Expansion}
\label{network_analysis}
\begin{figure}[ht]
\centering
\includegraphics[width=\columnwidth]{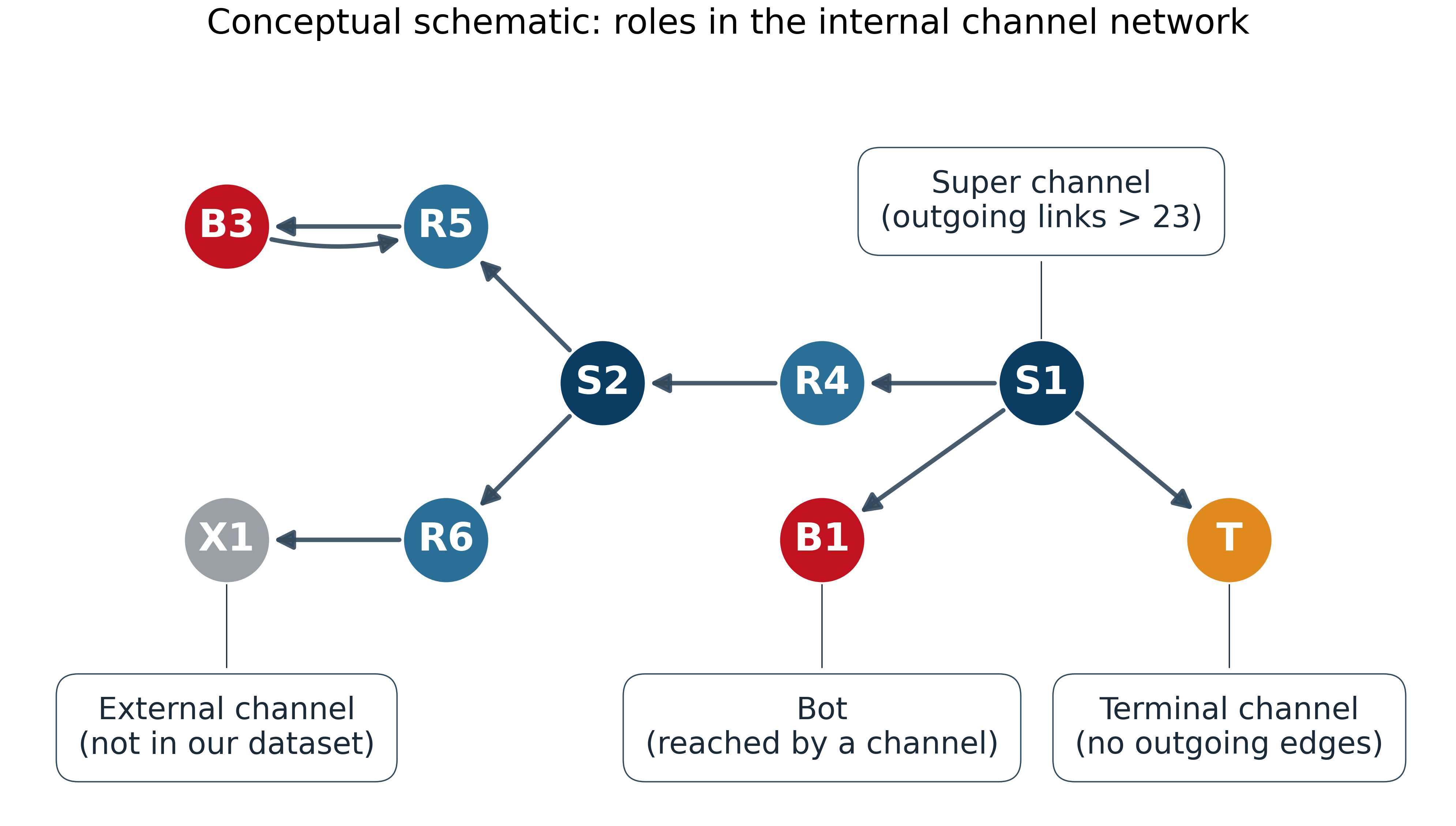}
\caption{Conceptual illustration of the network structure and semantic channel roles. Directed edges denote links between nodes. Dark blue nodes represent super channels, blue nodes regular channels, red nodes bots, orange nodes terminal channels, and gray nodes channels outside our dataset.}
\label{fig:concept_figure}
\end{figure}

In Section~\ref{qualitative_analysis}, we observed that seed channels frequently redirected users through intermediary Telegram channels and bots before exposing the actual pirated content. Here, we examine whether these patterns generalize at scale using our larger quantitative dataset. 
We model the internal promotion ecosystem as a directed graph in which each identifiable channel or bot represents a node, and each shared Telegram link forms a directed edge from the source entity to the promoted destination. This captures both channel-to-channel promotion and bot-mediated redirection within a unified structure. The resulting network contains 2,312 nodes (1,875 channels and 437 bots) and 3,843 directed edges. While the ecosystem is primarily channel-driven, bots form a secondary automation layer that supports content delivery and navigation. Of the mapped channels, only 648 originated from the source dataset, while 1,227 additional channels were discovered exclusively through internal redirection. This suggests that a substantial portion of the ecosystem remained hidden from traditional discovery approaches and only becomes visible through traversal of the promotion network itself. 
Connectivity within the graph was also dominated by channel-to-channel promotion, which accounts for 2,994 edges, followed by 583 channel-to-bot and 266 bot-to-channel edges. To interpret these structural patterns, we define three semantic channel roles based on their position in the network (Figure~\ref{fig:concept_figure}). \textit{Regular channels} share one or more internal Telegram links, \textit{super-channels} exhibit unusually high outward connectivity (more than 23 outgoing links based on the second standard deviation of the outgoing degree distribution ($\mu = 5.9554$, $\tilde{x} = 3$, $\sigma = 8.2207$), and \textit{terminal channels} receive incoming links without further redirecting users to additional internal destinations. 
Channels that exclusively promoted private invite links were excluded from outgoing edge analysis because such links cannot be reliably attributed to identifiable public destinations. Consequently, 64 invite-link-only channels were removed, although 9 still appeared in the source dataset because they were referenced through publicly observable links from other channels. 
The network exhibits \emph{strong interconnectedness}, with 93.92\% of channels participating in connected components of size two or greater, and only 6.08\% remaining isolated. This indicates that most piracy channels do not operate independently, but instead participate in densely connected promotion structures. 

To obtain a more granular view of these structures, we analyzed a highly active region within the largest connected component. This subgraph contained 35 channels connected through 212 directed channel-to-channel edges, with no bot participation, indicating a purely channel-driven routing structure. Connectivity within the subgraph was unevenly distributed, with some channels receiving links from up to 17 distinct sources while others receive links from only two. This concentration suggests that traffic is funneled toward a smaller subset of highly referenced destinations, while other channels function as intermediate or peripheral routing nodes.

Beyond direct promotion, the network also exhibits substantial transitive connectivity. We identified 177 indirect source-to-destination relationships involving 19 regular channels, where users can reach destinations through paths of length two or greater despite the absence of direct links. Together, the presence of both direct and indirect routing relationships reveals a decentralized multi-path architecture in which connectivity is distributed across several alternative routes rather than concentrated around a single critical node. As a result, users may continue reaching similar destinations even if individual channels are removed, suggesting that the ecosystem possesses substantial structural redundancy and resilience. 

At a broader scale, we identified 572 regular channels containing outgoing channel-to-channel links. Among them, 190 channels (32.04\%) redirected users to a single destination, while 344 (58.01\%) linked to multiple destinations, indicating that many channels distribute traffic across several downstream locations rather than functioning as isolated forwarding points. Within this structure, we identified 27 super-channels and 40 terminal channels. The super-channels act as large-scale routing hubs, with the most connected channel linking to 74 unique destinations and several others connecting to more than 40 neighboring channels. These highly connected hubs form coordinated sub-networks that further reinforce reachability and redundancy across the ecosystem. In contrast, terminal channels likely function as final content destinations where users consume material rather than continue navigating through additional redirection paths. 
Bots further extend this ecosystem by supporting automated delivery and redirection workflows. We identified 65 channels that linked exclusively to bots without directly promoting other channels.  Among the 437 identified bots, 192 promoted at least one internal URL, collectively generating 266 bot-to-channel connections. The largest bot-centered cluster contains 49 interconnected bots, while several channels were observed coordinating dozens of bots simultaneously for indexing, content delivery, download redirection, backup access, and automation. These patterns suggest that bots operate as an auxiliary infrastructure layer that increases scalability, redundancy, and persistence within the broader piracy ecosystem. 

%% file: sections/realworld_impact.tex
\section{Real-time framework}
\label{real_word_impact}
One of the primary motivations behind the taxonomy introduced in Section~\ref{taxonomy} is to enable automated identification of piracy-related posts and mapping of their underlying intent---information that can then be reported to relevant stakeholders to expedite the takedown of such content. In Section~\ref{quantitative_analysis}, we operationalize this using Gemma3:27B to categorize posts at scale. While effective, Gemma3:27B is a resource-intensive model that requires substantial GPU capacity, making it impractical for real-world deployment where stakeholders must process hundreds of thousands of data points every day, and such computational overhead can significantly slow down detection pipelines.

To address this limitation, we introduce two lightweight models finetuned on Llama3.2:1B (a significantly smaller and less memory hungry LLM), the first being a binary classifier that identifies whether a post constitutes piracy, and the second being a categorization model that then assigns one or more labels from the taxonomy to the detected post. We train them on a random sample of 100k posts from our large-scale dataset from Section~\ref{quantitative_analysis} that was labeled using Gemma3:27B. \jaishnoor{this is only for the categorization model tho?} We effectively distill the knowledge of a larger model into a much smaller one (Llama3.2:1B), a common approach for building cost-efficient detection systems using LLMs \cite{IBMKnowledgeDistillation}. To add benign samples to the training data, we randomly sampled 300k posts from 27.k channels using the Pushshift Telegram dataset~\cite{pushshifttelegram}, filtering out 12,746 empty posts.  
\\
\textbf{Training:}
Both the detection and categorization models were trained using an 80:20 train-test split, where 20\% of the dataset was held out for evaluation. While the detection model is evaluated on its output (piracy/benign), we evaluate the categorization model based on its cumulative accuracy in assigning a primary label and, where applicable, one or two secondary labels. Specifically, we report the accuracy of the primary label and the joint accuracy of both primary and secondary labels. This distinction is important, as a model may correctly predict the primary label while assigning an incorrect secondary label, or correctly assign secondary labels while predicting an incorrect primary label. Such partially correct predictions can introduce ambiguity and reduce the reliability of the output for downstream stakeholders. 
\begin{table}[t]
\centering
\caption{Performance of the categorization model under cumulative label settings. S(1) and S(2) denote the inclusion of one and two secondary labels, respectively.}
\label{tab:categorization_performance}
\resizebox{\columnwidth}{!}{%
\begin{tabular}{lcccc}
\hline
\textbf{Label Setting} & \textbf{Accuracy} & \textbf{Precision} & \textbf{Recall} & \textbf{F1} \\
\hline
Primary & 98.2\% & 98.6\% & 97.8\% & 98.2\% \\
Primary + S(1) & 95.6\% & 96.1\% & 94.8\% & 95.4\% \\
Primary + S(1) + S(2) & 94.1\% & 94.9\% & 93.2\% & 94.0\% \\
\hline
\end{tabular}%
}
\end{table}
The detection model showed an accuracy of 98\%, with an F1 score of 0.97. Table~\ref{tab:categorization_performance} highlights the performance of the categorization model across the cumulative labels. We observe that the model achieves strong performance when predicting the primary label alone (F1 = 98.2\%), indicating its ability to reliably capture the dominant intent of a post. When one secondary label is incorporated, performance decreases to an F1 of 95.4\%, and further to 94.0\% when two secondary labels are considered. This gradual decline reflects the increasing difficulty of fine-grained multi-label prediction. While primary labels capture the most salient signal, secondary labels introduce additional nuance and often overlap in semantics, making them harder to predict consistently. Despite this, the relatively small drop in performance demonstrates that the model retains strong capability in capturing secondary intent.

\subsection{The Anti-RIP pipeline}
\begin{figure}[ht]
\centering
  \includegraphics[width=0.9\columnwidth]{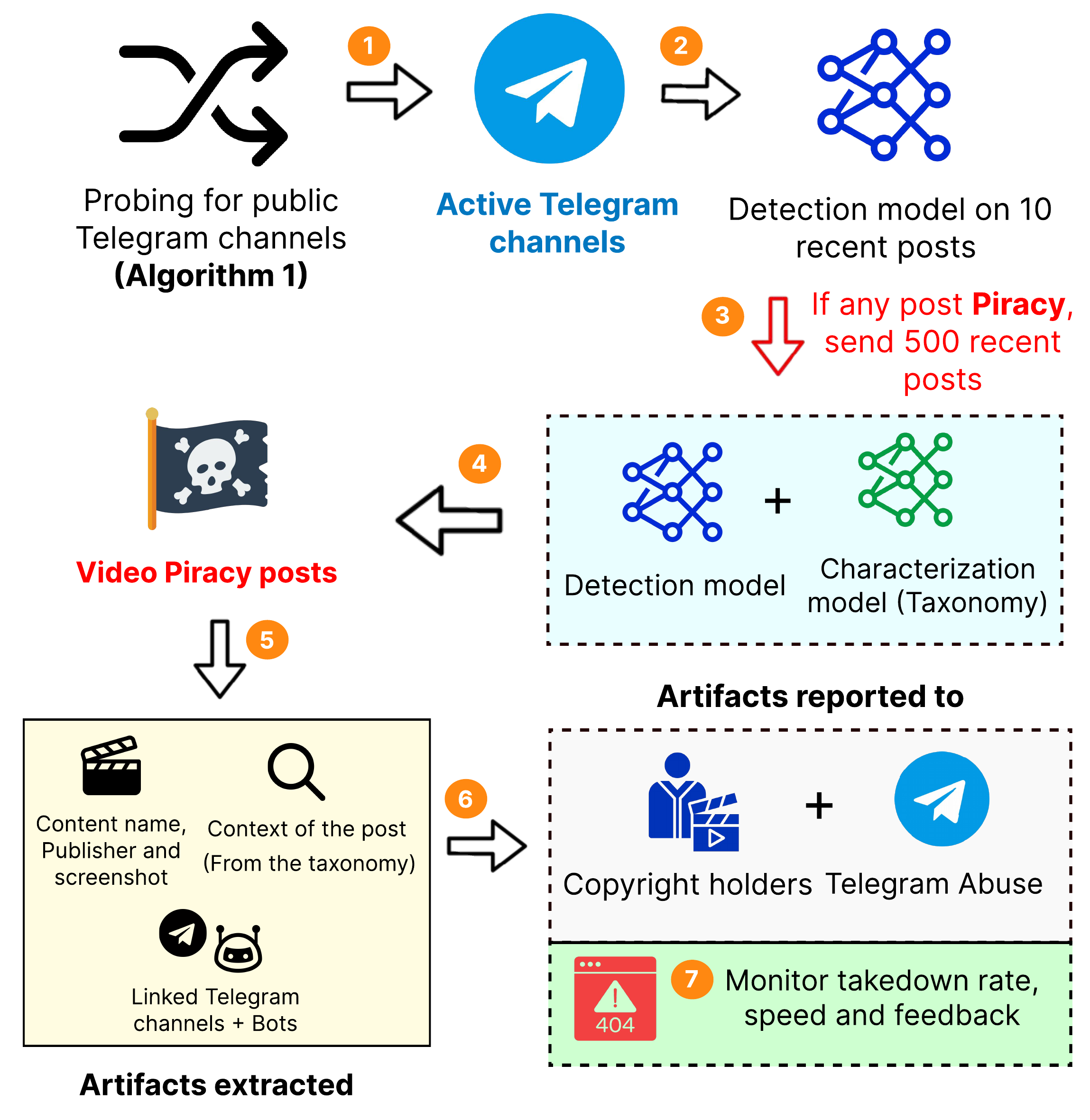}
 \caption{The Anti-RIP framework}
  \label{fig:anti_rip_framework}
\end{figure}

We operationalized our detection and categorization models as a real-time framework for identifying and reporting emerging piracy communities on Telegram. The framework consists of three stages: 1) continuous channel discovery, 2) automated detection and artifact enrichment, and 3) evidence-driven reporting to relevant stakeholders. Figure~\ref{fig:anti_rip_framework} presents an overview of the framework pipeline, while the following subsections detail each stage of its operation.
\newline
\textbf{Channel Discovery:} A central objective of the framework is to identify newly emerging piracy channels in near real time, enabling stakeholders to intervene before these communities accumulate substantial visibility and subscriber bases. Prior work~\cite{roy2025darkgram} further suggests that  enforcement actions are often more effective toward newly reported abuse communities than toward large, well-established ecosystems.
While the large-scale dataset in Section~\ref{quantitative_analysis} was collected through SearXNG, traditional search engines often exhibit delays in indexing newly created subdomains and pages~\cite{lefortier2013timely}, including Telegram channels.

Instead, we focus on probing for public Telegram channels whose usernames are openly exposed through \texttt{https://t.me/} by constructing a seed lexicon of 58 commonly occurring terms derived directly from the handles of channels previously discovered through SearXNG. We adopt this approach based on an empirical observation that many SearXNG-discovered channels, despite being collected independently, exhibited frequently recurring terms and naming patterns within their usernames - suggesting that video-piracy communities on the platform often follow common naming conventions. Using this vocabulary, we generate candidate usernames that satisfy Telegram's public handle constraints: 5 to 32 characters in length, composed of alphanumeric characters and underscores, and beginning with a letter. Candidate generation includes individual tokens, pairwise combinations (e.g., \texttt{w1\_w2}, \texttt{w1w2}), and optional higher-order compositions.
Algorithm~\ref{alg:candidate-generation} presents  the pseudocode for the candidate handle synthesis process. $\mathcal{L}$ denotes the seed lexicon and $\mathcal{H}$ denotes an optional set of observed Telegram-style handle strings corresponding to complete usernames extracted from previously observed Telegram links or historical crawl artifacts. For example, if historical data contains links such as \texttt{https://t.me/s/series\_rip\_hd} and \texttt{https://t.me/s/bollywood\_4k}, then $\mathcal{H}=\{\texttt{series\_rip\_hd},\texttt{bollywood\_4k}\}$. The algorithm splits each handle on underscores to recover fragments such as \emph{series}, \emph{rip}, \emph{hd}, \emph{bollywood}, and \texttt{4k}, which are added to the vocabulary and used to generate adjacent composites such as \texttt{series\_rip}, \texttt{seriesrip}, and \texttt{rip\_hd}, along with additional combinations involving the seed lexicon. If no historical handles are available, then $\mathcal{H}=\emptyset$, and candidate generation is driven solely by the seed vocabulary.
The algorithm constructs a normalized vocabulary $\mathcal{W}$ from these inputs and generates candidate handles through single-token fragments, adjacent handle-derived combinations, pairwise cross-products, concatenations, and optional higher-order compositions. The resulting candidate set $\mathcal{C}$ retains only strings that satisfy Telegram's public username constraints. Each candidate handle is subsequently converted into its corresponding public Telegram URL and probed to determine whether it resolves to an active public channel using the Telegram API.

To prioritize emerging communities, we further restrict discovery to channels whose earliest observed post was created within the previous seven days. In our Network analysis in Section~\ref{network_analysis}, we identified how piracy communities frequently exhibit interconnected and cross-promotional behavior- where channels actively advertise related channels, backup communities and bots.
Thus, for each active candidate channel, we retrieve up to 10 of its most recent posts and analyze their contents for references to additional Telegram channels, such as embedded \texttt{t.me} links. Newly observed links are recursively followed using a bounded breadth-first traversal up to depth 2, where depth 0 corresponds to channels/bots identified through handle synthesis, depth 1 represents channels directly referenced by those seeds, and depth 2 captures channels/bots discovered through subsequent inter-channel references. 
Using this approach, between February 3rd and April 10th, 2026, we monitored 249,133 newly discovered channels and 1,208 bots.
\begin{algorithm}[t]
\caption{Candidate Telegram handle generation for probing}
\label{alg:candidate-generation}
\begin{algorithmic}[1]
\Require Seed strings $\mathcal{L}$; optional handle strings $\mathcal{H}$
\Ensure Candidate handle set $\mathcal{C}$

\State $\mathcal{W}\gets \textsc{ValidFragments}(\mathcal{L}\cup \textsc{SplitHandles}(\mathcal{H}))$
\State $\mathcal{S}\gets \mathcal{W}$

\ForAll{$h\in\mathcal{H}$}
    \State $\mathcal{S}\gets \mathcal{S}\cup \textsc{AdjacentCombinations}(h)$
\EndFor

\ForAll{$(w_i,w_j)\in\mathcal{W}^2,\; w_i\neq w_j$}
    \State $\mathcal{S}\gets \mathcal{S}\cup\{w_i\texttt{\_}w_j,\; w_iw_j\}$
\EndFor

\State $\mathcal{S}\gets \mathcal{S}\cup \textsc{SampleHigherOrder}(\mathcal{W})$
\State $\mathcal{C}\gets\{s\in\mathcal{S}:\textsc{HandleOk}(s)\}$
\State \Return $\textsc{Shuffle}(\mathcal{C})$

\end{algorithmic}
\end{algorithm}
\newline
\textbf{Detection and Characterization:} Our detection framework initially analyzes the ten most recent posts collected from each newly discovered channel. If at least one of these posts is classified as video piracy, the framework performs a deeper analysis by retrieving up to 500 of the channel's most recent posts, including the initial ten, and evaluating them using the same detection model. For posts identified as piracy-related, the framework then attempts to infer both the distributed content and its associated production company by matching against the TMDb dataset, following a process similar to that described in Section~\ref{content_distribution_landscape}. The framework then collects additional supporting artifacts from the post metadata, including the screenshot of the posts, as well as both internal and external links. Finally, the post is automatically labeled by our Categorization model using the taxonomy (Section~\ref{taxonomy} to characterize the intent of the post. Using this overall workflow, we identified 802 piracy channels, which had a median age of 4.2 days and a median subscriber count of 521. These channels further exposed 299 interconnected channels through recursive network expansion, of which 237 were identified as regular intermediary channels, 58 as terminal channels, and 4 as super-channels. In addition, the framework discovered 108 bots. These linked channels had a much higher median age of 102 days and a median subscriber count of 3,855, suggesting that finding newer channels can help us identify older, more well-established channels as well. 
\\
\textbf{Reporting:} Using the artifacts and taxonomy-derived characteristics extracted from posts within channels identified as piracy, we generated comprehensive evidence-backed reports that were sent to Telegram's abuse email (abuse@telegram.org), as well as the impacted rights holders through their respective abuse emails. 
For the latter, we generated tailored reports for the affected rights holders based on the copyrighted material identified in each channel. For example, a report sent to a rights holder included the channel details, along with every post associated with that publisher’s contents, accompanied by the title of the distributed work, a screenshot of the post, extracted internal and external links, and the taxonomy-derived intent and behavioral characterization of the post. 
Overall, the 802 newly discovered piracy channels and 299 interconnected channels contained 58,391 unique posts classified as piracy and distributed 4,182 unique copyrighted titles associated with 104 rights holders. Due to organizational reporting constraints, notifications to content owners were limited to the 17 most prominent U.S.-based production companies, while all identified channel occurrences were reported to Telegram through its abuse reporting process. Of the 108 associated bots uncovered through recursive expansion, 53 were categorized as content delivery bots, 14 as dynamic retrieval bots, 39 as channel promotion bots, and only 2 as content ingestion bots.
\subsection{Impact of Reporting}
To assess the practical impact of our reporting efforts, we longitudinally tracked the status of the reported channels and bots from the time each entity was reported up to two weeks after the report was submitted.
\\
\textbf{Channels and bots:} During this period, 524 of the 1,101 reported channels and interconnected channels (47.6\%) became inaccessible, including 389 of the newly discovered piracy channels and 135 of the interconnected channels linked from them. 
To better understand whether these removals were attributable to our reporting, we also examined feedback received directly from Telegram. Of the 524 removed channels, Telegram explicitly confirmed enforcement takedown action for 309. For the remaining 215 channels, two possibilities exist: either the operators voluntarily removed the channels, or enforcement occurred without explicit feedback to us. However, given our observations in both the qualitative and quantitative analyses that piracy channels often persist for years with limited spontaneous disappearance, the latter explanation appears substantially more plausible. Moreover, because reports were also submitted to impacted rights holders, it is possible that downstream escalation by production companies contributed to enforcement outcomes for a portion of these channels, a phenomenon which is common for the removal of piracy forums and domains \cite{DBLP:journals/corr/abs-1804-02679}. 
Among the remaining 577 channels that were not fully removed, we observed substantial evidence of partial intervention at the content level. For 4,392 posts across 81 channels, Telegram explicitly notified us that action had been taken against specific posts; upon verification, these posts were no longer accessible and displayed Telegram-issued DMCA copyright warnings, as previously noticed in our qualitative and quantitative analysis in Sections~\ref{qualitative_analysis} and~\ref{quantitative_analysis} respectively. Motivated by this, we revisited reported posts for which no explicit feedback had been received and identified an additional 10,877 posts across 95 channels displaying the same warning, suggesting that rights holder escalation may also have contributed to these enforcement actions. Overall, these findings further confirm that even when Telegram does not disable an entire channel, it may still intervene by selectively removing infringing posts and issuing copyright enforcement notices, thereby disrupting distribution while leaving the broader channel intact.
On the other hand, we found that 71 of the 108 reported bots became inaccessible during this period. For 44 of these, Telegram explicitly confirmed enforcement action and indicated that the bots had been removed. For the remaining 27 inaccessible bots, a plausible explanation is that downstream escalation by rights holders contributed to their disruption. Given the central role bots play in content delivery and access control, and more importantly, the fact that they are often reused by multiple channels, their disruption can have positive cascading effect, simultaneously disrupting several interconnected channels and weakening the broader distribution infrastructure.

\textbf{Right-holders' response}:
Overall, we reported 14,742 piracy posts associated with copyrighted content to the abuse teams of 17 U.S.-based production houses, with reports generated and emailed on a per-occurrence basis whenever Anti-RIP identified content tied to a given publisher. 
For ethical and operational reasons, we do not disclose the identities of participating publishers or attribute outcomes to specific organizations, as doing so could expose enforcement practices or aid adversarial adaptation. Accordingly, we report only aggregate engagement and outcome statistics throughout this section.
Out of these publishers, 14 acknowledged receiving and reviewing our reports, while 9 explicitly stated that our reports contributed to their takedown or escalation efforts.

Interestingly, 5 of the 14 responding teams requested that reports be transitioned from event-driven submissions to 24-hour batched reporting, noting that receiving individual abuse notices for each detected infringement could overwhelm their analyst queues, whereas batched submissions allowed related infringements to be grouped into their coordinated internal enforcement workflows, particularly when multiple channels or mirror links were tied to the same content release. This suggests that, while real-time reporting supports rapid intervention, stakeholders may benefit from reporting modes aligned with their own internal operational workflows.

A notable qualitative insight from publisher responses was that contextual reporting often mattered as much as the infringement indicators themselves. In particular, 4 of the 14 responding teams explicitly acknowledged that the taxonomy-derived context in our reports improved their ability to assess and prioritize infringement notifications. This reinforces one of the central motivations behind our taxonomy-driven reporting design: richer abuse context can improve not only detection, but the effectiveness of intervention.

However, 3 of the 14 responding teams expressed a contrasting preference, requesting that future reports contain only the abuse URL or direct infringement artifact, without the additional contextual signals or supporting artifacts. This provided an opportunity to examine whether such preferences correlated with downstream enforcement outcomes. Focusing on channels for which Telegram did not explicitly confirm channel-level or post-level takedowns, we compared post-level DMCA enforcement actions associated with publishers that valued contextual reports versus those preferring URL-only reports. We found that publishers in the former group were associated with significantly higher rates of post-level DMCA removals on Telegram (median 31.4\% of reported posts per publisher) than the latter group (median 12.7\%), with the difference significant under a Mann-Whitney U test ($p<0.05$). While not establishing causality as there are several aspects that might factor into content takedown, this suggests that richer, contextual reporting may contribute to stronger downstream enforcement signals than artifact-only reporting alone.

%% file: sections/conclusion.tex
\section{Conclusion}
In this paper, we presented the first large-scale characterization of the video piracy ecosystem on Telegram, revealing a highly interconnected and resilient infrastructure built around channels, bots, and multi-stage distribution strategies. Using insights from our qualitative and quantitative analyses, we developed a fine-grained taxonomy of piracy-related behaviors and operationalized it through Anti-RIP, a real-time framework for identifying and reporting emerging piracy communities. Overall, our findings demonstrate both the scale and sophistication of Telegram-based video piracy, while highlighting the value of taxonomy-guided analysis for enabling more effective detection and intervention efforts.

%% file: refs.bib
@techreport{euipo2021socialmedia,
  author       = {{European Union Intellectual Property Office (EUIPO)}},
  title        = {Social Media---Discussion Paper: New and Existing Trends in Using Social Media for IP Infringement Activities and Good Practices to Address Them},
  institution  = {European Observatory on Infringements of Intellectual Property Rights},
  year         = {2021},
  month        = jun,
  url          = {https://euipo.europa.eu/tunnel-web/secure/webdav/guest/document_library/observatory/documents/reports/2021_Social_Media/2021_Social_Media_Discussion_Paper_FullR_en.pdf},
  note         = {Accessed: 2026-05-03}
}

@techreport{ustr2025notorious,
  author       = {{Office of the United States Trade Representative (USTR)}},
  title        = {2025 Review of Notorious Markets for Counterfeiting and Piracy},
  institution  = {Office of the United States Trade Representative},
  year         = {2026},
  month        = mar,
  url          = {https://ustr.gov/sites/default/files/files/Press/Releases/2026/2025%20Notorious%20Markets%20List%20%28final%29.pdf},
  note         = {Accessed: 2026-05-03}
}

@techreport{usco2020section512,
  author       = {{U.S. Copyright Office}},
  title        = {Section 512 of Title 17: A Report of the Register of Copyrights},
  institution  = {U.S. Copyright Office},
  year         = {2020},
  month        = may,
  url          = {https://www.copyright.gov/policy/section512/section-512-full-report.pdf},
  note         = {Published May 21, 2020; accessed 2026-05-03}
}

@article{kaye2021copyright,
  title={Copyright gossip: Exploring copyright opinions, theories, and strategies on {Y}ou{T}ube},
  author={Kaye, D Bondy Valdovinos and Gray, Joanne E},
  journal={Social Media+ Society},
  volume={7},
  number={3},
  pages={20563051211036940},
  year={2021},
  publisher={SAGE Publications Sage UK: London, England}
}

@inproceedings{zhang2022node,
  title={Node-imbalance learning on heterogeneous graph for pirated video website detection},
  author={Zhang, Shijun and Yin, Jiangyi and Li, Zhao and Yang, Rong and Du, Meijie and Li, Renjie},
  booktitle={2022 IEEE 25th international conference on computer supported cooperative work in design (CSCWD)},
  pages={834--840},
  year={2022},
  organization={IEEE}
}

@misc{google2016fights,
  key          = {Google},
  title        = {{How Google Fights Piracy}},
  year         = {2016},
  url          = {https://storage.googleapis.com/gweb-uniblog-publish-prod/documents/GO806_Google_FightsPiracy_eReader_final.pdf}
}

@article{danaher2020effect,
  author  = {Danaher, Brett and Hersh, Jonathan Samuel and Smith, Michael D. and Telang, Rahul},
  title   = {The Effect of Piracy Website Blocking on Consumer Behavior},
  year    = {2019},
  month   = aug,
  note    = {Available at SSRN: https://ssrn.com/abstract=2612063},
  doi     = {10.2139/ssrn.2612063}
}

@article{urban2017notice,
  title={Notice and takedown in everyday practice},
  author={Urban, Jennifer M and Karaganis, Joe and Schofield, Brianna},
  journal={UC Berkeley Public Law Research Paper},
  number={2755628},
  year={2017}
}

@inproceedings{lefortier2013timely,
  title={Timely crawling of high-quality ephemeral new content},
  author={Lefortier, Damien and Ostroumova, Liudmila and Samosvat, Egor and Serdyukov, Pavel},
  booktitle={Proceedings of the 22nd ACM international conference on Information \& Knowledge Management},
  pages={745--750},
  year={2013}
}

@inproceedings{li2022fighting,
  title={Fighting against piracy: An approach to detect pirated video websites enhanced by third-party services},
  author={Li, Zhao and Zhang, Shijun and Yin, Jiangyi and Du, Meijie and Zhang, Zhongyi and Liu, Qingyun},
  booktitle={2022 IEEE Symposium on Computers and Communications (ISCC)},
  pages={1--7},
  year={2022},
  organization={IEEE}
}

@article{danaher2014gone,
  title={Gone in 60 Seconds: The Impact of the {M}egaupload Shutdown on Movie Sales},
  author={Danaher, Brett and Smith, Michael D.},
  journal={International Journal of Industrial Organization},
  volume={33},
  pages={1--8},
  year={2014},
  doi={10.1016/j.ijindorg.2013.12.001},
  issn={0167-7187},
  publisher={Elsevier}
}

@dataset{pushshifttelegram,
  author       = {Baumgartner, Jason and Zannettou, Savvas and Blackburn, Jeremy and Cristofaro, Emiliano De},
  title        = {The {P}ushshift {T}elegram Dataset},
  year         = {2020},
  publisher    = {Pushshift},
  note         = {A large-scale dataset of public Telegram messages for research purposes},
}

@misc{telegram,
  author       = {{Telegram Messenger Inc.}},
  title        = {{Telegram Messenger}},
  howpublished = {\url{https://telegram.org}}
}

@article{mehta2025telegram1b,
  author       = {Ivan Mehta},
  title        = {Telegram founder {Pavel Durov} says app now has 1B users, calls {WhatsApp} a `cheap, watered down imitation'},
  journal      = {TechCrunch},
  year         = {2025},
  month        = mar,
  day          = {19},
  url          = {https://shorturl.at/S7211}
}

@inproceedings{roy2025darkgram,
  title={$\{$DarkGram$\}$: A $\{$Large-Scale$\}$ Analysis of Cybercriminal Activity Channels on Telegram},
  author={Roy, Sayak Saha and Vafa, Elham Pourabbas and Khanmohamaddi, Kobra and Nilizadeh, Shirin},
  booktitle={34th USENIX Security Symposium (USENIX Security 25)},
  pages={4839--4858},
  year={2025}
}

@misc{telemetr,
  title        = {Telemetr.io Global Catalog},
  author       = {{Telemetr.io}},
  year         = {2024},
  howpublished = {\url{https://telemetr.io}},
  note         = {Accessed: 2024-07-10}
}

@incollection{mckenzie202025,
  author    = {McKenzie, Jordi},
  title     = {Digital Piracy},
  booktitle = {Handbook of Cultural Economics, Third Edition},
  year      = {2020},
  publisher = {Edward Elgar Publishing},
  pages     = {228--234},
  month     = mar,
  doi       = {10.4337/9781788975803.00031}
}

@article{mooney2010napster,
  author  = {Mooney, Patrick and Samanta, Subarna and Zadeh, Ali H. M.},
  title   = {Napster and its Effects on the Music Industry: An Empirical Analysis},
  journal = {Journal of Social Sciences},
  year    = {2010},
  volume  = {6},
  number  = {3},
  pages   = {303--309},
  issn    = {1549-3652}
}

@article{alexander2002peer,
  title={Peer-to-peer file sharing: The case of the music recording industry},
  author={Alexander, Peter},
  journal={Review of Industrial Organization},
  volume={20},
  number={2},
  pages={151-161},
  year={2002},
  journal = {Review of Industrial Organization},
doi = {10.1023/A:1013819218792}
}

@online{fedsSeizeDomainNames2010,
 author  = {News Desk},
 date    = {2010-06-30},
 title   = {Feds Seize Domain Names of Movie Piracy Websites},
 organozation = {Courthouse News Service},
 url     = {https://www.courthousenews.com/feds-seize-domain-names-of-movie-piracy-websites/},
 urldate = {2026-04-21}
}

@online{usassistsbulgaria,
 author  = {{Office of Public Affairs}},
 date    = {2026-01-30},
 title   = {{U.S.} Law Enforcement Assists {B}ulgarian Law Enforcement in Taking Down Three of the Largest Piracy Sites in The {European Union}},
 organozation = { U.S. Department of Justice},
 url     = {https://shorturl.at/rpqjG},
 urldate = {2026-04-21}
}

@online{takedown123movues,
 author  = {Waqas},
 date    = {2018-04-03},
 title   = {Streaming site 123movies has been shut down},
 organization = {HackRead},
 url     = {https://hackread.com/streaming-site-123movies-shut-down-and-its-alterntives/},
 urldate = {2026-04-21}
}

@article{piratebay2014disruption,
 author  = {Yevgeniy Sverdlik},
 date    = {2014-12-10},
 title   = {{Pirate Bay} Down After {S}tockholm Area Data Center Raid},
 journal = {Data Center Knowledge},
 url     = {https://www.datacenterknowledge.com/business/pirate-bay-down-after-stockholm-area-data-center-raid},
 urldate = {2026-04-21}
}

@inproceedings{oest2019phishfarm,
  title={Phishfarm: A scalable framework for measuring the effectiveness of evasion techniques against browser phishing blacklists},
  author={Oest, Adam and Safaei, Yeganeh and Doup{\'e}, Adam and Ahn, Gail-Joon and Wardman, Brad and Tyers, Kevin},
  booktitle={2019 IEEE Symposium on Security and Privacy (SP)},
  pages={1344--1361},
  year={2019},
  organization={IEEE}
}

@inproceedings{lauinger2013clickonomics,
  title={Clickonomics: Determining the Effect of Anti-Piracy Measures for One-Click Hosting.},
  author={Lauinger, Tobias and Szydlowski, Martin and Onarlioglu, Kaan and Wondracek, Gilbert and Kirda, Engin and Kruegel, Christopher},
  booktitle={NDSS},
  year={2013}
}

@inproceedings{oest2020sunrise,
  title={Sunrise to sunset: Analyzing the end-to-end life cycle and effectiveness of phishing attacks at scale},
  author={Oest, Adam and Zhang, Penghui and Wardman, Brad and Nunes, Eric and Burgis, Jakub and Zand, Ali and Thomas, Kurt and Doup{\'e}, Adam and Ahn, Gail-Joon},
  booktitle={29th $\{$USENIX$\}$ Security Symposium ($\{$USENIX$\}$ Security 20)},
  year={2020}
}

@inproceedings{bitaab2020scam,
  title={Scam pandemic: How attackers exploit public fear through phishing},
  author={Bitaab, Marzieh and Cho, Haehyun and Oest, Adam and Zhang, Penghui and Sun, Zhibo and Pourmohamad, Rana and Kim, Doowon and Bao, Tiffany and Wang, Ruoyu and Shoshitaishvili, Yan and others},
  booktitle={2020 APWG Symposium on Electronic Crime Research (eCrime)},
  pages={1--10},
  year={2020},
  organization={IEEE}
}

@article{ntd,
  author    = {Perel, Maayan and Elkin-Koren, Niva},
  title     = {Accountability in Algorithmic Copyright Enforcement},
  journal   = {Stanford Technology Law Review},
  year      = {2016},
  volume    = {19},
  number    = {3},
  pages     = {473--532},
}

@article{dmca,
  author    = {Bar-Ziv, Sharon and Elkin-Koren, Niva},
  title     = {Behind the Scenes of Online Copyright Enforcement: Empirical Evidence on Notice \& Takedown},
  journal   = {Connecticut Law Review},
  year      = {2018},
  volume    = {50},
  number    = {2},
  pages     = {339--386},
}

@article{robotakedown,
  author    = {Carpou, Zoe},
  title     = {Robots, Pirates, and the Rise of the Automated Takedown Regime: Using the {DMCA} to Fight Piracy and Protect End-Users},
  journal   = {Columbia Journal of Law \& the Arts},
  year      = {2016},
  volume    = {39},
  number    = {4},
  pages     = {551--590},
}

@report{ustr2025review,
  author       = {{Office of the United States Trade Representative}},
  title        = {2025 Review of Notorious Markets for Counterfeiting and Piracy},
  year         = {2026},
  institution  = {Office of the United States Trade Representative},
  type         = {Report},
  url          = {https://ustr.gov/sites/default/files/files/Press/Releases/2026/2025%20Notorious%20Markets%20List%20(final).pdf},
}

@online{harris2015reboot,
  author       = {Harris, Donald P.},
  title        = {Time to Reboot? {Rethinking the Digital Millennium Copyright Act}},
  year         = {2015},
  organization = {Temple University Beasley School of Law},
  url          = {https://law.temple.edu/10q/time-to-reboot-rethinking-the-digital-millennium-copyright-act/},
  note         = {faculty article in temple's business law zine)},
}

@article{whacamole_ohagan2024,
  author  = {O'Hagan, Cassie},
  title   = {Indie Authors' Battle Against Piracy on Digital Publishing Platforms: An Endless Game of Whac-A-Mole},
  journal = {Michigan State Law Review},
  year    = {2024},
  volume  = {2024.2},
  pages   = {501},
}

@online{stout2025siteblocking,
  author       = {Stout, Kristian and Morris, Julian and Ramakrishnan, Subi},
  title        = {Site Blocking and Incentive-Compatible Solutions to Illicit Online Activity},
  year         = {2025},
  organization = {International Center for Law \& Economics},
  url          = {https://laweconcenter.org/resources/site-blocking-and-incentive-compatible-solutions-to-illicit-online-activity/},
  note         = {Law \& Economics Center resource},
}

@misc{doj2026hackerforum,
  author       = {{U.S. Department of Justice, Office of Public Affairs}},
  title        = {{United States} Leads Dismantlement of One of the World's Largest Hacker Forums},
  year         = {2026},
  month        = mar,
  day          = {4},
  url          = {https://www.justice.gov/opa/pr/united-states-leads-dismantlement-one-worlds-largest-hacker-forums},
  note         = {Press release, Criminal Division / FBI Cybercrime operation},
}

@inproceedings{dhanwate2025empowering,
  author={Dhanwate, Gaurav and Jadhav, Archana and Lavate, Indrajit and Bawage, Samarth and Aghavane, Rushikesh and Patil, Dipali},
  booktitle={2025 IEEE International Conference on Blockchain and Distributed Systems Security (ICBDS)}, 
  title={Empowering Digital Rights: Real-Time Detection of Piracy Channels on {T}elegram Using Machine Learning and Network Analysis}, 
  year={2025},
  volume={},
  number={},
  pages={1-6},
  doi={10.1109/ICBDS67396.2025.11379699}}

@article{maddusila2024copyright,
  author  = {Maddusila, Sitti Fatimah and Fithrah and Hendrawan, Deni and Adiman and Purnamasari, Andi Intan and Supriyadi},
  title   = {Copyright Restrictions in Social Media Markets: A Legal Enforcement Challenge},
  journal = {International Journal of Criminal Justice Sciences},
  year    = {2024},
  volume  = {19},
  number  = {2},
  pages   = {},
  note    = {July-December 2024},
}

@article{bal2023telegrampiracy,
author = {Bal, Meghna},
year = {2023},
month = {05},
pages = {1-15},
title = {Audio-visual piracy on {T}elegram: a perspective on monetization models, pirate strategies and industrial pathways},
volume = {31},
journal = {Contemporary South Asia},
doi = {10.1080/09584935.2023.2204220}
}

@misc{lymishchenko,
author = {Lymishchenko, Valeriia and Kamar, Eden and Botchkovar, Ekaterina and Maimon, David},
year = {2025},
month = {01},
pages = {},
title = {Comparative Analysis of Cyber Fraud Ecosystems: Telegram and Dark Web Platforms in Digital Criminal Landscapes},
doi = {10.2139/ssrn.5722747}
}

@article{morgia2021darktelegram,
  author  = {La Morgia, Massimo and Mei, Alessandro and Mongardini, Alberto Maria and Wu, Jie},
  title   = {Uncovering the Dark Side of {T}elegram: Fakes, Clones, Scams, and Conspiracy Movements},
  journal = {arXiv preprint arXiv:2111.13530},
  year    = {2021},
  doi     = {10.48550/arXiv.2111.13530},
  url     = {https://arxiv.org/abs/2111.13530},
}

@TechReport{POUWELSE2008701,
type={19th ITS Biennial Conference, Bangkok 2012: Moving Forward with Future Technologies - Opening a Platform for All},
institution={International Telecommunications Society (ITS)},
author={Potgieter, Petrus H.},
title={High-definition content and file sharing networks},
year={2012},
number={72518},
doi={None},
url={https://ideas.repec.org/p/zbw/itsb12/72518.html},
}

@online{AFP2016JapanPiracy,
  author       = {Stacey Leasca},
  title        = {Japan introduces strict anti-piracy laws},
  date         = {2016-07-31},
  organization = {GlobalPost},
  url          = {https://theworld.org/stories/2016/07/31/japan-introduces-strict-anti-piracy-laws},
  note         = {Published by The World (PRX)}
}

@online{JustiaPiracyEntertainment,
  title        = {Piracy in the Entertainment Industry \& Legal Penalties},
  organization = {Justia},
  url          = {https://www.justia.com/entertainment-law/piracy-in-the-entertainment-industry/},
  note         = {Accessed 2026-05-06}
}

@inproceedings{fbgraphsearch,
  author={Khan, Zubeida Casmod and Mashiane, Thulani},
  booktitle={2014 Information Security for South Africa}, 
  title={An analysis of {F}acebook's graph search}, 
  year={2014},
  volume={},
  number={},
  pages={1-8},
  doi={10.1109/ISSA.2014.6950517}}

@online{BlockX2024TelegramPiracyPlatform,
  author       = {{Block X}},
  title        = {Why {T}elegram is the Go-To Platform for Pirates},
  date         = {2025-06-06},
  url          = {https://www.linkedin.com/pulse/why-telegram-go-to-platform-pirates-blockxtechs-tpovc/},
  organization = {Block X},
  note         = {Published on LinkedIn Pulse; accessed 2026-05-06}
}

@online{DMCAForceTelegramDMCA,
  author       = {{DMCA Force}},
  title        = {How to Use {Telegram DMCA} for Copyright Infringement},
  date         = {2025-06-04},
  url          = {https://dmcaforce.com/how-to-use-telegram-dmca-for-copyright-infringement/},
  organization = {DMCA Force},
  note         = {Accessed 2026-05-06}
}

@online{ICC2011CounterfeitingPiracyImpacts,
  author       = {{International Chamber of Commerce (ICC)}},
  title        = {The Economic Impacts of Counterfeiting and Piracy},
  date         = {2017-02-03},
  url          = {https://iccwbo.org/news-publications/policies-reports/economic-impacts-counterfeiting-piracy-report-prepared-bascap-inta/},
  organization = {International Chamber of Commerce (ICC)},
  note         = { accessed 2026-05-06}
}

@online{NCTA2019PiracyEconomyImpact,
  author       = {{NCTA---The Internet \& Television Association}},
  title        = {How Digital Piracy is Harming {A}merica’s Economy},
  date         = {2019-09-11},
  url          = {https://www.ncta.com/news/how-digital-piracy-is-harming-americas-economy},
  organization = {NCTA},
  note         = {Accessed 2026-05-06}
}

@online{IBMKnowledgeDistillation,
  author       = {Dave Bergmann}, 
  title        = {What is knowledge distillation?},
  url          = {https://www.ibm.com/think/topics/knowledge-distillation},
  organization = {IBM},
  note         = {Accessed 2026-05-06}
}

@misc{FBIIC32023Report,
  author       = {{Federal Bureau of Investigation Internet Crime Complaint Center (IC3)}},
  title        = {2023 Internet Crime Report},
  year         = {2024},
  howpublished = {\url{https://www.ic3.gov/annualreport/reports/2023_ic3report.pdf}},
  note         = { accessed 2026-05-06}
}

@misc{Ashwin2020Telegram2GB,
  author       = {Ashwin Karthik},
  title        = {Telegram increases file size limit to {2GB}, adds support for multiple accounts on Desktop, profile videos on mobile},
  year         = {2020},
  month        = jul,
  day          = {27},
  howpublished = {\url{https://shorturl.at/V1Av1}},
  note         = {gHacks Technology News; accessed 2026-05-06}
}

@misc{Knibbs2015RapidShareShutdown,
  author       = {Kate Knibbs},
  title        = {Another One of the Most Popular File-Sharing Sites Is Getting Shut Down},
  year         = {2015},
  month        = feb,
  day          = {10},
  howpublished = {\url{https://gizmodo.com/another-one-of-the-most-popular-file-sharing-sites-is-g-1684967540}},
  note         = { accessed 2026-05-06}
}

@misc{DigitalWebSolutions2024TelegramStats,
  author       = {Raghav Tayal}, 
  title        = {Telegram Users Statistics: Growth, Demographics, and Insights},
  year         = {2025},
  month     = may,
  day          = {26} , 
  howpublished = {\url{https://www.digitalwebsolutions.com/blog/telegram-users-statistics/}},
  note         = {Accessed 2026-05-06}
}

@misc{Kumar2026iPhoneAndroidUsers,
  author       = {Naveen Kumar},
  title        = {i{P}hone vs {A}ndroid Users Market Share (New Stats 2026)},
  year         = {2026},
  month        = mar,
  day          = {12},
  howpublished = {\url{https://www.demandsage.com/iphone-vs-android-users/}},
  note         = {DemandSage; accessed 2026-05-06}
}

@inproceedings{stayinalive,
  author    = {Tina Marjanov and Taro Tsuchiya and Konstantinos Ioannidis and Jack Hughes and Nicolas Christin and Alice Hutchings},
  title     = {Stayin' Alive: How Global Stolen Data Markets Thrive on {T}elegram},
  booktitle = {USENIX Security Symposium (USENIX Security '26)},
  year      = {2026},
  url       = {https://www.usenix.org/system/files/conference/usenixsecurity26/sec26_prepub_marjanov.pdf}, 
  note      = {Accessed 2026-05-06}
}

@article{Fu_Teo_Seng_2012, 
title={The bandwagon effect on participation in and use of a social networking site}, 
volume={17}, 
url={https://firstmonday.org/ojs/index.php/fm/article/view/3971}, 
DOI={10.5210/fm.v17i5.3971}, 
 number={5}, 
 journal={First Monday}, 
 author={Fu, W. Wayne and Teo, Jaelen and Seng, Seraphina},
 year={2012}, 
 month={May} }

@article{Katona2011DiffusionSocialNetwork,
  author  = {Zsolt Katona and Peter P. Zubcsek and Miklos Sarvary},
  title   = {Network Effects and Personal Influences: The Diffusion of an Online Social Network},
  journal = {Journal of Marketing Research},
  year    = {2011},
  volume  = {48},
  number  = {3},
  doi     = {10.1509/jmkr.48.3.425},
  url     = {https://journals.sagepub.com/doi/full/10.1509/jmkr.48.3.425}
}

@misc{mishra2020tacklingonlineabusesurvey,
      title={Tackling Online Abuse: A Survey of Automated Abuse Detection Methods}, 
      author={Pushkar Mishra and Helen Yannakoudakis and Ekaterina Shutova},
      year={2020},
      eprint={1908.06024},
      archivePrefix={arXiv},
      primaryClass={cs.CL},
      url={https://arxiv.org/abs/1908.06024}, 
}

@misc{AppleSupport108047,
  author       = {{Apple Inc.}},
  title        = {i{C}loud+ plans and pricing},
  year         = {2026},
  howpublished = {\url{https://support.apple.com/en-us/108047}},
  note         = {Apple Support; accessed 2026-05-06}
}

@misc{GoogleOneSupportStorage,
  author       = {{Google}},
  title        = {How your {G}oogle storage works}, 

  howpublished = {\url{https://support.google.com/googleone/answer/9312312}},
  note         = {Google One Help; accessed 2026-05-06} 
}

@misc{terabox,
  title        = {1024{GB} Cloud Storage},
  howpublished = {\url{https://www.terabox.com/}},
}

@misc{terashare,
  title        = {Sharing large files has never been easier},
  howpublished = {\url{https://terashare.net/en/}},
}

@misc{gofile,
  author       = {{GoFile}},
  title        = {{GoFile}---Cloud Storage Made Simple},
  howpublished = {\url{https://gofile.io/}},
}

@article{CHAUDHRY201777,
title = {The looming shadow of illicit trade on the internet},
journal = {Business Horizons},
volume = {60},
number = {1},
pages = {77-89},
year = {2017},
issn = {0007-6813},
doi = {https://doi.org/10.1016/j.bushor.2016.09.002},
url = {https://www.sciencedirect.com/science/article/pii/S0007681316300908},
author = {Peggy E. Chaudhry},
}

@article{Marx2013StorageWars,
  author  = {Marx, Nick},
  title   = {Storage Wars: Clouds, Cyberlockers, and Media Piracy in the Digital Economy},
  journal = {The Journal of e-Media Studies},
  year    = {2013},
  volume  = {3},
  number  = {1},
  article = {4},
  doi     = {10.1349/PS1.1938-6060.A.426},
  url     = {https://digitalcommons.dartmouth.edu/joems/vol3/iss1/4}
}

@phdthesis{Makitalo2024TrackingStolenVideo,
  author       = {M{\"a}kitalo, Mikko},
  title        = {Tracking the stolen video product on the internet: focus on illicit streaming of copyrighted content},
  year         = {2024},
  school       = {H{\"a}me University of Applied Sciences},
  url          = {https://urn.fi/URN:NBN:fi:amk-2024052113744},
  note         = {URN: URN:NBN:fi:amk-2024052113744; accessed 2026-05-06}
}

@article{DBLP:journals/corr/abs-1804-02679,
  author       = {Damilola Ibosiola and
                  Benjamin A. Steer and
                  {\'{A}}lvaro Garc{\'{\i}}a{-}Recuero and
                  Gianluca Stringhini and
                  Steve Uhlig and
                  Gareth Tyson},
  title        = {{Movie Pirates of the Caribbean}: Exploring Illegal Streaming Cyberlockers},
  journal      = {CoRR},
  volume       = {abs/1804.02679},
  year         = {2018},
  url          = {http://arxiv.org/abs/1804.02679},
  eprinttype   = {arXiv},
  eprint       = {1804.02679},
  bibsource    = {dblp computer science bibliography, https://dblp.org}
}

@misc{Lynch2025GeoRestrictedStreaming,
  author       = {Gina Lynch},
  title        = {The Ultimate List of Geo-restricted Streaming Services},
  year         = {2026},
  month        = apr,
  day          = {30},
  howpublished = {\url{https://secureblitz.com/list-of-geo-restricted-streaming-services/}},
  note         = {SecureBlitz; accessed 2026-05-06}
}

@misc{Ibeh2025NetflixNigeriaAffordability,
  author       = {Royal Ibeh},
  title        = {Equal prices, unequal earnings: Netflix affordability still skewed against {N}igerians},
  year         = {2025},
  month        = oct,
  day          = {24},
  howpublished = {\url{https://tinyurl.com/4ptjd7tk}},
  note         = {BusinessDay Nigeria; accessed 2026-05-06}
}

@misc{Broderick2025StreamingPiracyReturn,
  author       = {Gabriel V Rindborg},
  title        = {Can’t pay, won’t pay: impoverished streaming services are driving viewers back to piracy},
  year         = {2025},
  month        = aug,
  day          = {14},
  howpublished = {\url{https://shorturl.at/fVRg3}},
  note         = {The Guardian; accessed 2026-05-06}
}

@article{LuRajaviDinner2021OTTpiracy,
  author  = {Lu, Shijie and Rajavi, Koushyar and Dinner, Isaac},
  title   = {The Effect of Over-the-Top Media Services on Piracy Search: Evidence from a Natural Experiment},
  journal = {Marketing Science},
  year    = {2021},
  volume  = {40},
  number  = {3},
  pages   = {548--568},
  doi     = {10.1287/mksc.2020.1256},
  url     = {https://pubsonline.informs.org/doi/10.1287/mksc.2020.1256}
}

@misc{Boulifi2026PiracyProblemStreaming,
  author       = {Tharwa Boulifi},
  title        = {The Piracy Problem Streaming Platforms Can’t Solve},
  year         = {2026},
  month        = mar,
  howpublished = {\url{https://www.wired.com/story/the-piracy-problem-streaming-platforms-cant-solve/}},
  note         = {WIRED; accessed 2026-05-06}
}

@misc{Lewkowicz2025WhenBotsGetSmarter,
  author       = {Natalie Lewkowicz},
  title        = {When Bots Get Smarter: Static Fraud Tools, No Match for {AI}},
  year         = {2025},
  month        = nov,
  day          = {3},
  howpublished = {\url{https://www.darwinium.com/resources/the-evolution-blog/when-bots-get-smarter}},
  note         = {Darwinium Evolution Blog; accessed 2026-05-06}
}

@manual{searxng,
  title        = {Welcome to {SearXNG}},
  author       = {{SearXNG Project}},
  year         = {2026},
  url          = {https://docs.searxng.org/},
  note         = {Accessed: 2026-05-07}
}

@mastersthesis{WhyPay,
    author = {Theodore Giletti},
    title = {Why Pay If It’s Free? {S}treaming, Downloading, and Digital Music Consumption in the `i{T}unes Era'},
    school = {London School of Economics and Political Science},
    year = {2012}
}

@misc{tmdb,
  author       = {{The Movie Database (TMDB)}},
  title        = {Getting Started},
  year         = {2026},
  url          = {https://developer.themoviedb.org/docs/getting-started},
  note         = {Accessed: 2026-05-07}
}
